\documentclass[11pt,a4paper]{article}
\usepackage{latexsym,amssymb,amsmath,amsthm,amsfonts,enumerate,verbatim,xspace,
exscale}

\input xy
\xyoption{all} \CompileMatrices \UseComputerModernTips

\usepackage{graphicx,tabularx}
\usepackage{amsmath,amsbsy,amsfonts,amssymb}
\usepackage{color}
\usepackage{amsthm}
\usepackage{mathtools}
\usepackage{lineno}

\makeatletter
\newcommand{\boldparagraph}[1]{\par\vspace{1.75ex \@plus.3ex \@minus.1ex}\noindent\textbf{#1}\hspace{.6em}}
\makeatother
\title{The Optimal Reduction of a 2-Body Problem in $\mathbb{R}^3$}

\date{}

\author{{\sc{Hern\'an Cendra}\thanks{
         Departamento de Matem\'{a}tica, Universidad Nacional del Sur, Av.\ Alem 1253, 
       8000 Bah\'{\i}a Blanca, Argentina.}
        }  \ \
\author{M. E. Garc\'ia  \footnotemark[2]}
        {\sc{Mar\'ia Eugenia Garc\'ia}\thanks{CMaLP - Centro de Matem\'atica de La Plata, Universidad Nacional de La Plata,  Calles 50 y 115, La Plata, Argentina.}
\thanks{Facultad de Ciencias Exactas, Universidad Nacional de La Plata, Calles 50 y 115, La Plata, Argentina. } } \ \
 }
\newtheorem{theorem}{Theorem}[section]
\newtheorem{corollary}[theorem]{Corollary}

\newtheorem{proposition}[theorem]{Proposition}

\theoremstyle{definition}

\begin{document}




\maketitle
\begin{abstract}
In this paper we describe optimal reduction for the system of two bodies in $\mathbb{R}^3$  whose Hamiltonian is invariant under rotations and translations. In doing this, we introduce parametrizations and charts which help giving explicit expressions in order to deal with  geometric and dynamical aspects of the reduction process. For this system, the standard assumptions of the Marsden-Weinstein reduction process  are only partially satisfied while optimal reduction can be readily applied, and we study a comparison between those two reduction processes. We describe potential applications to the study of Post-Newtonian Hamiltonian systems for binary systems in astronomy.

\end{abstract}

{
\setcounter{tocdepth}{2}
\tableofcontents
}
\section{Introduction}

Symmetry is one of the fundamental notions in physics and mathematics. It belongs to the very notion of a physical object since it represents the invariance under the change of coordinates or, in a more physical language, the observer independence. On the other hand, conservation laws, in many cases, are directly related to some group of symmetries. For instance, the Euclidean group $SE(3)$ of translations and rotations in $\mathbb{R}^3$ is directly related to the fundamental conservation laws of linear and angular momentum.
These geometric ideas in mechanics have evolved for a long time, some of the important references are \cite{AM}, \cite{Arnold2}, \cite{MDL89}, \cite{Goldstein}, \cite{Hamilton},  \cite{Jacobi}, \cite{Saletan}, \cite{Lagrange}, \cite{MR}, \cite{MW-coments}, \cite{Poincare1}, \cite{Poincare2} and references therein. They are presently  a very active field of research from the purely mathematical point of view and also from the point of view of physics, engineering and many other fields, pure or applied. 

An important point is that the presence of symmetry allows to perform \textit{reduction} (\cite{AM}, \cite{LREPE}, \cite{LRBS}, \cite{HRBS}, \cite{MR-reduction}, \cite{MR}, \cite{O-R}, \cite{O-R_libro}). For part of the history of reduction see ~\cite{MW-coments}.

There are several reduction theories and each one of them assumes specific hypothesis. The Marsden-Weinstein reduction uses the fundamental notion of momentum map (\cite{AM}, \cite{guillsternbrg}, \cite{guillsternbrg3}, \cite{guillsternbrg2}, \cite{guillsternbrg4}, \cite{Smale1}, \cite{Smale2}) and has been applied in many important examples. However, the assumptions of this theory, including  the very existence of the momentum map, are not satisfied in several cases of interest, see for instance \cite{AM}, \cite{Smale1}, \cite{Smale2}. In particular, we will see that the Marsden-Weinstein reduction theory can be applied only partially to the system studied in this paper. 

Recent fundamental works in this direction in which the notions of \textsl{optimal momentum map} and \textsl{optimal reduction} are introduced, have been developed in \cite{O-R}, ~\cite{O-R_libro}.

In this paper, which is devoted only to the 2-Body Problem (2BP), the full generality of the results in \cite{O-R},~\cite{O-R_libro} is not needed. In fact, we only use the case of  a symplectic manifold with a symplectic, proper and globally Hamiltonian group action which includes, by definition, existence and equivariance of a momentum map. In this case, as it was proven in \cite{O-R},~\cite{O-R_libro}, there is a useful formula for calculating those level sets in given specific cases, see formula (5.5.5) in ~\cite{O-R_libro}. This formula is one of the important tools from the optimal momentum theory that we use in the paper.
\ \\

The 2BP has been studied mathematically and generalized in several directions since the time of Newton (\cite{AM}, \cite{Arnold2}), including the case of a not necessarily Newtonian potential, like the so called Post-Newtonian Approximations which take into account relativistic effects.

The main purpose of the paper is to describe, for the 2BP with $SE(3)$ symmetry, certain manifolds and maps, like the level sets of the optimal momentum map and their optimal reduced spaces with the corresponding bundle projection. We do this by using physically meaningful parametrizations where the parametrization of the reduced spaces is related to the original position and momentum variables in a simple and explicit way. So we deliver a concrete and not just an abstract characterization of the optimal reduced spaces and the bundle projections.  
\ \\

This paper is organized as follows. In Section \ref{section2} we recall basic notions about the optimal momentum map and optimal reduction and, also, some basic facts about the 2BP. In Section \ref{section 3} we calculate the level sets of the optimal momentum map, their optimal reduced spaces and the corresponding principal bundle projection. In order to do this we divide the problem in $10$ cases, some of which are conjugate to some others, as we will explain.
In Section \ref{sec:prel} we summarize the results and present some of the conclusions conveniently in the form of a Table. This Table helps to understand the reduction process and the relationship between optimal reduction and Marsden-Weinstein reduction. Finally we discuss a potential application to binary systems in astronomy. 

\section{Background}\label{section2}

In the next subsection we will recall some basic facts about optimal reduction from \cite{O-R} and ~\cite{O-R_libro}.

\subsection{Optimal momentum map and optimal reduction for proper, symplectic and globally Hamiltonian actions}\label{subsection 2.1}

Let $(M,\omega)$ be a symplectic manifold and let a left action of a Lie group $G$ on $M$ which is symplectic, proper and globally Hamiltonian. Let $C^{\infty}(M)^G$ be the set of $G$-invariant functions on $M$. The {\bf\emph{G-characteristic distribution}} $E$ is, by definition, the smooth generalized distribution on $M$ spanned by the set of Hamiltonian vector fields $X_f$ with $f\in C^{\infty}(M)^G$. Define $G_E$ as being the pseudogroup generated by the local flows of $X_f$ with $f\in C^{\infty}(M)^G$. 

In Proposition 3.3 of \cite{O-R} the fact that the $G$-characteristic distribution $E$ is smooth, completely integrable and its maximal integral manifolds are submanifolds which are the orbits of the pseudogroup $G_E$ is proven. The {\bf \emph{optimal momentum map}} is the canonical projection $\mathcal{J}:M\rightarrow M/G_E$, where $M/G_E$ is, by definition, the {\bf \emph{momentum space}}.

\ \\{\bf Notation.} For $x\in M$ we will denote $\mathcal{J}(x)\subset M$ the orbit of the action of $G_E$ on $x$. On the other hand we will normally denote $\rho$ an element of the momentum space $M/G_E$. Consequently we will sometimes denote using a slight abuse of notation, $\rho_x=\mathcal{J}(x)$, or simply $\rho=\mathcal{J}(x)$ or $\mathcal{J}(x)=\mathcal{J}^{-1}(\rho)$. 
\\

For a given subgroup $H$ of $G$, let $M_H$ be the {\bf \emph{isotropy type submanifold}} given by $M_H=\{z\in M:\ G_z=H\}$ where, using a standard notation, $G_z$ is the isotropy subgroup of the action of $G$ on $M$. The following results were proven in \cite{O-R} and \cite{O-R_libro}, in particular formula (\ref{formula optimal momentum}) was proven as formula (5.5.5) of \cite{O-R_libro}. 
\ \\

\textit{Let} $(M,\omega)$ \textit{be a symplectic manifold and $G$ be a Lie group acting on $M$ in a symplectic, proper and globally Hamiltonian fashion with associated equivariant momentum map} $J:M\rightarrow \mathfrak{g}^*$. \textit{If $E$ is the $G$-characteristic distribution, then for any $m\in M$} \begin{equation}\label{formula distribucion}E(m)=Ker T_mJ\cap T_mM_{G_m}.
\end{equation}

\textit{Moreover, the $G_E$-orbit of the point $m$}, $\mathcal{J}^{-1}(\rho)$ with $\rho=\mathcal{J}(m)$, is the connected component containing $m$ of the submanifold $J^{-1}(\mu)\cap M_{G_m}$, that is 
\begin{equation}\label{formula optimal momentum}
\mathcal{J}^{-1}(\rho)=(J^{-1}(\mu)\cap M_{G_m})_{c.c.m} 
\end{equation}
where $\mu=J(m)\in\mathfrak{g}^*$.

The following corollary from the previous facts is important for the present paper, see \cite{O-R} sections 4.2 and 4.3.

\begin{corollary}\label{corolario formula} 
In the case in which some non empty set $J^{-1}(\mu)\cap M_{G_m}$ is connected, using formula (\ref{formula optimal momentum}) we can deduce that $\mathcal{J}^{-1}(\rho)=J^{-1}(\mu)\cap M_{G_m}$.

If every non empty set $J^{-1}(\mu)\cap M_{G_m}$ is connected we can deduce that all values of the optimal momentum can be biunivocally labelled $\rho_{(\mu,S)}$ where $\mu$ represents a value of the momentum map and $S$ represents an isotropy subgroup such that $J^{-1}(\mu)\cap M_S\neq \emptyset$. Then $\mathcal{J}^{-1}(\rho_{(\mu,S)})= J^{-1}(\mu)\cap M_S$, for any $(\mu, S)$ such that $J^{-1}(\mu)\cap M_S$ is nonempty. This establishes an injective map $\Xi:M/G_E\longrightarrow J(M)\times {\bf Is}$, where ${\bf Is}$ is the set of all isotropy subgroups. In particular, for each $\mu \in J(M)$ we have the decomposition \begin{equation}\label{j como union de optimos}J^{-1}(\mu)=\displaystyle\bigcup_{(\mu,S)\in\, Im(\Xi)}\mathcal{J}^{-1}(\rho_{(\mu,S)}).\end{equation}
\end{corollary} 

The situation described in Corollary \ref{corolario formula} is related to the Connectedness Hypothesis, see Chapter 8 of \cite{O-R_libro}, and it occurs in the case of the 2BP.

\boldparagraph{Equivariance of the optimal momentum map and isotropy subgroup $G_{\rho}$.} 
Let $x \in M$, $\rho= \mathcal{J} (x)$ and $g \in G$. Then the following property holds, see \cite{O-R} or ~\cite{O-R_libro} section 5.5.13,
\begin{equation}\label{equivariance optimal}
g\mathcal{J}(x) = \mathcal{J}(gx)
\end{equation}
or, equivalently, using the previous notational convention
$g \mathcal{J}^{-1}\left(\rho_x\right) = \mathcal{J}^{-1}\left(\rho_{gx}\right)$.

Here we are using the notation $gZ := \{gx\ |\ x\in Z\}$, where $Z$ is any subset of $M$.

This observation helps in many cases since, instead of applying formula (\ref{formula optimal momentum}) to calculate the optimal momentum map at $gx$ we can apply formula  (\ref{equivariance optimal}) if we know the optimal momentum map at $x$. 
\ \\

The action of $G$ on the momentum space $M/G_E$ is defined by $g\rho=\mathcal{J}(gx)$, where $\rho=\mathcal{J}(x)$. The fact that this action is well defined is equivalent to  (\ref{equivariance optimal}). For a given $\rho\in M/G_E$, we define $G_{\rho}$ as being the isotropy subgroup of $\rho$ under this action. Using formula (\ref{equivariance optimal}) it follows that the condition $g \in G_\rho$ is equivalent to $g \mathcal{J}^{-1}\left(\rho_x\right) = \mathcal{J}^{-1}\left(\rho_{x}\right)$ or $\mathcal{J}(x)=g\mathcal{J}(x)$.

\boldparagraph{Optimal reduction.} In the following we collect some fundamental results that we need in the present paper especially in Theorem \ref{Teorema principal}. These results have been proven in a more general context in \cite{O-R}, sections 4.2 and 4.3. See also Section 9.1.7 in \cite{O-R_libro}. 
\ \\

\textit{Let $(M,\omega)$ be a symplectic manifold and $G$ a Lie group acting properly, symplectically and in a globally Hamiltonian fashion on $M$ with associated equivariant momentum map $J:M\rightarrow \mathfrak{g}^*$ and optimal momentum map $\mathcal{J}:M\rightarrow M/G_E$. Then, for any $\rho\in M/G_E$ the following assertions hold.}

\begin{enumerate} 
\item \textit{The isotropy subgroup $G_{\rho}$ acts on the submanifold $\mathcal{J}^{-1}(\rho)$, and the corresponding orbit space $M_{\rho}:=\mathcal{J}^{-1}(\rho)/G_{\rho}$ can be endowed with a unique smooth structure that makes the canonical projection $\pi_{\rho}:\mathcal{J}^{-1}(\rho)\rightarrow \mathcal{J}^{-1}(\rho)/G_{\rho}$ a submersion. The quotient space $M_{\rho}$ endowed with this smooth structure is called the {\bf \emph{optimal reduced phase space}}.}

\item \textit{The optimal reduced space $M_{\rho}=\mathcal{J}^{-1}(\rho)/G_{\rho}$ has a unique symplectic structure $\omega_{\rho}$ characterized by $$\pi^*_{\rho}\omega_{\rho}=i^*_{\rho}\omega,$$ where $\pi_{\rho}:\mathcal{J}^{-1}(\rho)\rightarrow M_{\rho}$ is the canonical projection and $i_{\rho}:\mathcal{J}^{-1}(\rho)\rightarrow M$ is the inclusion.}
\item \textit{Given an invariant Hamiltonian $H$ on $M$, the  restriction $H\vert_{\mathcal{J}^{-1}(\rho)}=i^*_{\rho}H$ defines uniquely a reduced Hamiltonian $H_{\rho}$ on $M_{\rho}$ that satisfies the condition $\pi^*_{\rho}H_{\rho}=i^*_{\rho}H$. For any given initial condition $m_0\in \mathcal{J}^{-1}(\rho)$, a solution curve $m(t)$ with initial condition $m_0$ passes to a quotient curve $\pi_{\rho}(m(t))$ which is the solution of $H_{\rho}$ with initial condition $\pi_{\rho}(m_0)$.}   
\end{enumerate}
\vspace{.2cm}

In the next subsection we collect some facts about the symmetry and the momentum map for the 2BP, see \cite{AM}.

\subsection{Configuration space, symmetry group and momentum map for the 2BP}\label{subsection2.2}

We will use the description of \cite{AM} for the Euclidean group $SE(3)$ of rotations and translations $(A,a)$ of $\mathbb{R}^3$, as the semidirect product $SE(3)=SO(3)\circledS{\mathbb{R}}^{3}$. We will call $\mathcal{T}  = \{(I, a): a \in \mathbb{R}^3\}$ the subgroup of translations, which is an abelian and normal subgroup. For given $v_0 \in  \mathbb{R}^3 - \{0\}$  we will call $\mathcal{T}_{v_0}  = \{(I, \xi v_0):\ \xi\in \mathbb{R}\}$ the subgroup of translations parallel to  $v_0$. 

Recall that given groups $G$, $N$ and a group homomorphism $\varphi:G \rightarrow Aut(N)$ the semidirect product $G\circledS_{\varphi} N$, or simply $G\circledS N$, is given by the following product on $G \times N$, $(g_1, n_1).(g_2, n_2)=(g_1 g_2, n_1 \varphi (g_1)(n_2))$. Two cases that we use in this paper are the following. First, 
$SO(3) \circledS \mathbb{R}^3$, where $\varphi(A)(x) = Ax$. Second, let $N\subseteq G$ be a normal subgroup such that $G = N \cup \{b\} $ with $b^2 = e$, so $\{e , b\}$ is a subgroup isomorphic to $\mathbb{Z}_2$,  and $\varphi$ be defined by conjugation, namely, $\varphi (b) (n) = bnb$. Then one can check that $G$ is isomorphic to $\{e , b\} \circledS N$, where the isomorphism is given by $(b, n) \rightarrow nb$. 

The configuration space of the 2BP is $Q=\mathbb{R}^3\times \mathbb{R}^3$ and the action $\phi$ of $SE(3)$ on $Q$ is the diagonal action given by $(A,a)(q^{1},q^{2})=(Aq^{1}+a,Aq^{2}+a)$. The phase space is $M=T^*Q$ with the canonical symplectic form $\Omega_M$ and the action of the group on $M$ is the cotangent lift of $\phi$.

Using the identification $(\mathbb{R}^3)^*\simeq\mathbb{R}^3$ given by the canonical inner product on $\mathbb{R}^3$, we have an identification $T^*Q\simeq\mathbb{R}^3\times\mathbb{R}^3\times\mathbb{R}^3\times\mathbb{R}^3$ and the cotangent lift action $T^*\phi$ is given by  $(A,a)(q^1,q^2,p^1,p^2)=(Aq^1+a,Aq^2+a,Ap^1,Ap^2)$. This action is symplectic, proper and globally Hamiltonian and the momentum map is given by $J(q,p)=(\alpha,u)$ where  
\begin{align}
\label{preimagen del momento alpha}\alpha&=q^1\times p^1+q^2\times p^2\\
\label{preimagen del momento u}u&=p^1+p^2.
\end{align}

Here we have used the identification $\mathfrak{se}(3)^*=(\mathfrak{so}(3))^*\times(\mathbb{R}^3)^*\cong\mathbb{R}^3\times\mathbb{R}^3$. Using this identification, the coadjoint action of $SE(3)$ on $\mathfrak{se}(3)^*$ is given by $Ad^*_{(A,a)^{-1}}(\alpha,u)=(A\alpha-Au\times a,Au)$.

\section{Optimal reduction for the 2BP with $SE(3)$-invariant Hamiltonian}\label{section 3}

In this section we are going to apply known fundamental facts proven in \cite{O-R} and \cite{O-R_libro} and Corollary \ref{corolario formula} to the case of the 2BP. The main results of the paper are in Theorem \ref{Teorema principal}. 

\subsection{Isotropy subgroups of the action $T^*\phi$ and their isotropy type manifolds $M_{SE(3)_{(q,p)}}$} \label{3.1}

Let $(q,p)=(q^1,q^2,p^1,p^2)\in\mathbb{R}^3\times\mathbb{R}^3\times\mathbb{R}^3\times\mathbb{R}^3$, then using the definition we obtain that
$$SE(3)_{(q,p)}=\left\{(A,a)\in SE(3):Aq^1+a=q^1,\ Aq^2+a=q^2,Ap^1=p^1,\ Ap^2=p^2\right\}.$$

We will introduce a useful notation which helps to describe the class $\bold{Is}$ of all isotropy subgroups of the action $T^*\phi$. For $x,\ y\in\mathbb{R}^3$ we define the following subgroups, where each one of them, as we will show in Proposition \ref{MH},  is an element of $\bold{Is}$ and conversely.

$$G^1(x)=\{(A,a)\in SE(3): a=x-Ax\}\simeq SO(3).$$

For $y\neq 0$, $$G^2(y)=\{(A,a)\in
SE(3):Ay=y, a=0\}\simeq SO(2).$$

For $x\times y\neq 0$, 
$$G^3(x,y)=\{(A,a)\in SE(3):Ay=y, a=x-Ax\}\simeq SO(2).$$

Finally we define $$G^4=\{(I,0)\}.$$

\boldparagraph{Parametrizations of classes of isotropy subgroups.} We have defined four classes of subgroups of $SE(3)$, namely, $G^1$, $G^2$, $G^3$, $G^4$ which are parametrized as follows:

\begin{itemize}\renewcommand{\labelitemi}{$-$}
\item $G^1$ is parametrized by $x\in \mathbb{R}^3$ and the parametrization is bijective.
\item $G^2$ is parametrized by $y\in \mathbb{R}^3-\{0\}$ and the parametrization is bijective modulo scale factors, that is
$G^2(y_1) = G^2(y_2)$ iff there exists $\lambda \neq 0$ such that $y_2 = \lambda y_1$. We will denote $[y]$ the equivalence class under scale, then $G^2 ([y]) := G^2 (y)$ is well defined and $G^2$ is bijectively parametrized by $[y]$. The set $P^2$ of classes $[y]$ is naturally identified with $\mathbb{R}P^2$, namely, for each $[y]$ there is a unique line $\ell_{[y]}$ containing $y$. 
\item $G^3$ is parametrized by $x, y \in \mathbb{R}^3-\{0\}$ satisfying $x\times y \neq 0$ and the parametrization is bijective modulo 
certain transformations, namely, $G^3(x_1, y_1) = G^3(x_2, y_2)$ if and only if there exist $\mu$, $\nu \neq 0$, 
such that $(x_2, y_2) = (x_1 + \mu y_1,  \nu y_1)$. We will denote $[x, y]$ the equivalence class under the previous transformations, then $G^3([x, y]):=G^3(x, y)$ is well defined and $G^3$ is bijectively parametrized by $[x, y]$. The set of classes $[x,y]$ is naturally identified with a fiber bundle $P^3$ with base $\mathbb{R}P^2$. For each  $\ell_{[y]}$ in $\mathbb{R}P^2$ the fiber $P^3_{[y]}$ is the plane perpendicular to $\ell_{[y]}$ which contains $0\in \mathbb{R}^3$,  minus $\{0\}$.
In fact, for each $[x,y]$ there is a uniquely determined $\ell_{[y]}$ and a uniquely determined $x\in P^3_{[y]}$.

\item $G^4$ has only one element namely $(I,0)$. 
\end{itemize}

Besides, we have:
\begin{itemize}\renewcommand{\labelitemi}{$-$}
\item For each $(x,y)$ there is an isomorphism between $G^2(y)$ and $G^3(x,y)$ given by conjugation, $G^3(x,y)=(I,x)G^2(y)(I,-x)$.
\item No subgroup in the class $G^2$ is equal to a subgroup in the class $G^3$; no subgroup in the class $G^1$ is isomorphic to a subgroup in the class $G^2$, $G^3$ or $G^4$; no subgroup in the class $G^4$ is isomorphic to a subgroup in the class $G^2$ or $G^3$. 
\item We have the disjoint union $G^1\cup G^2\cup G^3\cup G^4$. \item Note that each one of the subsets $G^1$, $G^4$ and $G^2\cup G^3$ is invariant by conjugation by any element $(A,a)\in SE(3)$.
\end{itemize}

We shall omit the proof of the previous assertions, which is not difficult.

\begin{proposition}\label{MH}
\begin{enumerate}[(a)]

\item The equality $\bold{Is}=G^1\cup G^2\cup G^3\cup G^4$ holds.
\item All isotropy type submanifolds are $M_{G^1(x)}, M_{G^2(y)}, M_{G^3(x,y)}, M_{G^4}$ where: 

\begin{enumerate}[$(b_1)$]
\item For $x\in
\mathbb{R}^3$, $M_{G^1(x)}=\{(x,x,0,0)\}$.
\item For $y\in\mathbb{R}^3$ with $y\neq 0$ we have the disjoint union decomposition
\begin{equation}\label{MG2}
M_{G^2(y)}=\displaystyle\bigcup_{i=1}^{4}B_i(y),\end{equation}
where $B_1(y)=\{(\lambda y,\beta y,0,0):\lambda,\beta\in\mathbb{R}, \lambda\neq\beta\}$; $B_2(y)=\{(\lambda y,\beta y,0,\delta y):\lambda,\beta,\delta\in\mathbb{R}, \delta\neq 0\}$; $B_3(y)=\{(\lambda y,\beta y,\gamma y,0):\lambda,\beta,\gamma\in\mathbb{R},\gamma\neq 0\}$ and $B_4(4)=\{(\lambda y,\beta y,\gamma y,\delta y):\lambda,\beta,\gamma,\delta\in\mathbb{R},\gamma,\delta\neq 0\}$.

\item For $x,y\in\mathbb{R}^3$ such that $x\times y\neq 0$, we have the disjoint union decomposition
\begin{equation}\label{MG3}
M_{G^3(x,y)}=\displaystyle\bigcup_{i=1}^{4}D_i(x,y),
\end{equation}
where $D_1(x,y)=\{(\lambda y+x,\beta y+x,0,0):\lambda,\beta\in\mathbb{R}, \lambda\neq\beta\}$; $D_2(x,y)=\{(\lambda y+x,\beta y+x,0,\delta y):\lambda,\beta,\delta\in\mathbb{R}, \delta\neq 0\}$; $D_3(x,y)=\{(\lambda y+x,\beta y+x,\gamma y,0):\lambda,\beta,\gamma\in\mathbb{R}, \gamma\neq 0\}$ and $D_4(x,y)=\{(\lambda y+x,\beta y+x,\gamma y,\delta y):\lambda,\beta,\gamma, \delta\in\mathbb{R}, \gamma\neq 0, \delta\neq 0\}$.

\item $M_{G^4}$ is a disjoint union given by
\begin{equation}\label{MG4}
M_{G^4}=\displaystyle\bigcup^{4}_{i=1}M_i,
\end{equation}
where $M_1=\{(q^1,q^2,0,p^2):q^1-q^2 \nparallel p^2\}$; $M_2=\{(q^1,q^2,p^1,0):q^1-q^2 \nparallel p^1\}$; $M_3=\{(q^1,q^2,p^1,p^2):p^1\parallel p^2,\ p^1\neq 0,\ p^2\neq 0,\ q^1-q^2\nparallel p^1\}$ and $M_4=\{(q^1,q^2,p^1,p^2):p^1\nparallel p^2\}$.
\end{enumerate}
\end{enumerate}
\end{proposition}

\ \\\emph{Proof of (a)} 

By inspection of \ref{Apendice A} we can deduce that the disjoint union $G^1 \cup G^2\cup G^3\cup G^4$ equals ${\bf Is}$. 

\ \\\emph{Proof of (b)}

This proof can be deduced essentially by inspection of \ref{Apendice A} and $(a)$. More precisely, we can proceed as follows.

To prove $(b_1)$ we can show by inspection of \ref{Apendice A} that if $m=(x,q^2,p^1,p^2)$ is such that
$SE(3)_m=SE(3)_{(x,x,0,0)}=G^1(x)$ then $p^1=p^2=0$ and $q^2=x$.

Now we will  prove $(b_2)$. We can see from ($a_2$) that the points
$m=(\lambda y,\beta y,0,0)$ with
$\lambda,\beta\in\mathbb{R}$ and $\lambda\neq\beta$ satisfy $SE(3)_m=SE(3)_{m_{2}}= G^2(y)$.

Given $m=(\lambda y,\beta y,0,\gamma y)$ with
$\lambda,\beta,\gamma\in\mathbb{R}$, $\gamma\neq 0$, we can see from ($a_6$) that $SE(3)_m=SE(3)_{m_{6}}=G^2(y)$.

Let $m=(\lambda y,\beta y,\gamma y,0)$ with
$\lambda,\beta,\gamma\in\mathbb{R}$, $\gamma\neq 0$, we can see from ($a_9$) that $SE(3)_m=SE(3)_{m_{9}}=G^2(y)$.

Finally given $m=(\lambda y,\beta y,\gamma y,\delta y)$ with
$\lambda,\beta,\gamma,\delta\in\mathbb{R}$, $\gamma,\delta\neq 0$, we can see from ($a_{12}$) that $SE(3)_m=SE(3)_{m_{12}}=G^{2}(y)$.

To finish the proof of $(b_2)$ we only need to note that ($a_2$), ($a_6$), ($a_9$), ($a_{12}$), are the only cases of \ref{Apendice A} where a group in the class $G^2$ appears.

Now we will  prove $(b_3)$. We can see from ($a_3$) that the points
$m=(\lambda y+x,\beta y+x,0,0)$ with
$\lambda,\beta\in\mathbb{R}$ and $\lambda\neq\beta$ satisfy  $SE(3)_m=SE(3)_{m_{3}}=G^3(x,y)$.

Let $m=(\lambda y+x,\beta y+x,0,\delta y)$ with
$\lambda,\beta,\delta\in\mathbb{R}$ and $\delta\neq 0$, then we can see from ($a_5$) that $SE(3)_m=SE(3)_{m_{5}}=G^3(x,y)$.

Given $m=(\lambda y+x,\beta y+x,\gamma y,0)$ with
$\lambda,\beta,\gamma\in\mathbb{R}$ and $\gamma\neq 0$, we can see from ($a_8$) that $SE(3)_m=SE(3)_{m_{8}}=G^3(x,y)$.

Finally for $m=(\lambda y+x,\beta y+x,\gamma y,\delta y)$ with $\gamma,\delta$ not zero, we can see from ($a_{11}$) that $SE(3)_{m}=SE(3)_{m_{11}}=G^3(x,y)$.

To finish the proof of $(b_3)$ we only need to note that ($a_3$), ($a_5$), ($a_8$) and ($a_{11}$), are the only cases of \ref{Apendice A} where a group in the class $G^3$ appears.

Finally, to  prove $(b_4)$ we can see that $M_{G^4}$ is a disjoint union corresponding to ($a_4$), ($a_7$), ($a_{10}$) and ($a_{13}$).
\ \\

In Theorem \ref{Teorema principal} we are going to consider  $10$ cases which include exactly all the non empty intersections $J^{-1}(\alpha_0,u_0)\cap 
M_{SE(3)_{(q,p)}}$. Note that $J^{-1}(0,0)\cap M_{G^4}=\emptyset$. First we shall consider \emph{Cases 1, 2, 3, 4, 5} in which one of the following conditions is satisfied: $(q_0,p_0)\in M_{G^1(x)}$, for some $x\in\mathbb{R}^3$; $(q_0,p_0)\in M_{G^2(y)}$, for some $y\in\mathbb{R}^3-\{0\}$; $(q_0,p_0)\in M_{G^3(x,y)}$, for some $x,y\in\mathbb{R}^3$ with $x\times y\neq 0$. Then we shall consider the \emph{Cases 6, 7, 8, 9} and \emph{10} in which the  condition $(q_0,p_0)\in M_{G^4}$ is satisfied. We will prove that in each one of the 10 cases the  set $J^{-1}(\alpha_0,u_0)\cap M_{SE(3)_{(q,p)}}$ is connected and therefore we will be able to apply Corollary \ref{corolario formula}. So we will be able to conclude that each such intersection corresponds biunivocally to a value of the optimal momentum map.

As we will show,  \emph{Cases 4, 5, 9, 10} are conjugate, in the sense of formula (\ref{equivariance optimal}), to \emph{Cases 2, 3, 7, 8} respectively, which simplifies some proofs. \emph{Cases 1} and \emph{6} do not have a conjugate case. There is a particular relationship between \emph{Cases 7} and \emph{8} since the latter is essentially a restriction of the former to the chart $W_p$ defined in \ref{Apendice B}.

\begin{theorem}\label{Teorema principal}
\ \\\begin{enumerate}[(A)]

\item The assertions stated in the following cases are satisfied

\boldparagraph{Case 1.} Let $(q_0,p_0)\in M_{G^1(x_0)}$,
$(q_0,p_0)=(x_0,x_0,0,0)$ with $x_0\in \mathbb{R}^3$, then $J(q_0,p_0)=(0,0)$. 

Let $\rho^1_{(0,0;x_0)}:=\mathcal{J}(q_0,p_0)$, then

 $\mathcal{J}^{-1}(\rho^1_{(0,0;x_0)})=J^{-1}(0,0)\cap M_{G^1(x_0)}=\{(x_0,x_0,0,0)\}$,

 $G_{\rho^1_{(0,0;x_0)}}=G^1(x_0)$,

 $M_{\rho^1_{(0,0;x_0)}}=\{[(x_0,x_0,0,0)]\},$
with $[(x_0,x_0,0,0)]=\{(x_0,x_0,0,0)\}$.

 The reduced symplectic form is trivial.

\boldparagraph{Case 2.} Let $(q_0,p_0)\in
M_{G^2(y_0)}$, with
$J(q_0,p_0)=(0,0)$, $y_0\in\mathbb{R}^3-\{0\}$ and $\rho^2_{(0,0; y_0)}:=\mathcal{J}(q_0,p_0)$, then
\begin{equation}\label{Caso 2}
\begin{array}{rcl}
\mathcal{J}^{-1}(\rho^2_{(0,0; y_0)})&=&J^{-1}(0,0)\cap M_{G^2(y_0)}\vspace{.2cm}\\
&=&\{(\lambda y_0,\beta y_0,0,0):\lambda,\beta\in\mathbb{R}, \lambda\neq\beta\}\vspace{.2cm}\\
&\cup&\{(\lambda y_0,\beta y_0,\gamma y_0,-\gamma y_0):\lambda,\beta,\gamma\in\mathbb{R},
\gamma\neq 0\}\simeq\mathbb{R}^3-\{\ell\},
\end{array}
\end{equation}
where $\ell$ is the straight line given by $\gamma=0$ and $\lambda-\beta=0$, which is connected so Corollary \ref{corolario formula} can be applied and the first equality of (\ref{Caso 2}) holds true. This gives a parametrization of $\mathcal{J}^{-1}(\rho^2_{(0,0; y_0)})$ in terms of the parameters 
$(\lambda, \delta,\gamma)$, where $\delta=\lambda-\beta$, in the domain $\gamma\neq 0$ or $\delta \neq 0$. The presymplectic form on $\mathcal{J}^{-1}(\rho^2_ {(0,0;y_0)})$ becomes $|y_0|^2 d\delta \wedge d\gamma$.

 $G_{\rho^2_{(0,0; y_0)}}=\{(A,a)\in SE(3):Ay_0=\pm y_0\
\mbox{and}\ a=\eta y_0, \eta\in\mathbb{R}\}$.

The subgroup $G^+_{\rho^2_{(0,0;y_0)}}= G^2(y_0) \circledS \mathcal{T}_{y_0}$ is a normal subgroup. If $B\in SO(3)$ satisfies $B^2=I$ and $By_0=-y_0$ then   
$G^+_{\rho^2_ {(0,0;y_0)}} \cup \{(B, 0)\}$ generates $G_{\rho^2_{(0,0;y_0)}}$. Moreover, since the group $\mathbb{Z}^{(2)}_2 :=\{(I,0), (B,0)\}$ acts by conjugation on $G^+_{\rho^2_{(0,0;y_0)}}$, the group $G_{\rho^2_ {(0,0;y_0)}}$ is naturally isomorphic to the semidirect product $\mathbb{Z}^{(2)}_2 \circledS G^+_{\rho^2_{(0,0;y_0)}}$. 

The map $\pi_{\rho^2_{(0,0;y_0)}}$ can be obtained in two stages \cite{O-R_libro}.  Using the parametrization, in the first stage we have the bundle $\pi^+_{\rho^2_{(0,0;y_0)}}: \mathcal{J}^{-1} (\rho^2_{(0,0;y_0)})\rightarrow M^+_{\rho^2_{(0,0;y_0)}}$, where
 $M^+_{\rho^2_{(0,0;y_0)}}:=\mathcal{J}^{-1} (\rho^2_{(0,0;y_0)}) /G^+_{\rho^2_{(0,0;y_0)}}\simeq
\mathbb{R}^2 - \{(0,0)\}$, with coordinates
$(x^+, y^+) = (\delta, \gamma)$,  or we can use polar coordinates
$x^+= r^+cos(\varphi^+), y^+= r^+sin(\varphi^+)$.

The group $G^{+}_{\rho^2_{(0,0;y_0)}}$ contains $\mathcal{T}_{y_0}$ as a normal subgroup whose action is free and  $G^2(y_0)$ as a subgroup whose action is trivial. These actions commute and $\pi^{+}_{\rho^2_{(0,0;y_0)}}$
is a trivial principal bundle isomorphic to
$\mathcal{T}_{y_0}\times M^{+}_{\rho^2_{(0,0;y_0)}}$.
A section $\sigma^+_{\rho^2_{(0,0;y_0)}}$ of $\pi^+_{\rho^2_{(0,0;y_0)}}$ giving this trivialization is
\begin{equation}\label{sigma2}
\sigma^+_{\rho^2_{(0,0;y_0)}}(x^+, y^+)=(x^+ y_0,0y_0, y^+ y_0, -y^+ y_0).
\end{equation}
In the second stage, since the action of the group $\mathbb{Z}^{(2)}_2$ on $\mathcal{J}^{-1} (\rho^2_ {(0,0;y_0)})$ passes to the quotient $M^+_{\rho^2_ {(0,0;y_0)}}$, we obtain a double covering $D^2:M^+_{\rho^2_ {(0,0;y_0)}} \rightarrow M^+_{\rho^2_ {(0,0;y_0)}}/ \mathbb{Z}^{(2)}_2$. Then we obtain $M_{\rho^2_{(0,0;y_0)}}=M^+_{\rho^2_ {(0,0;y_0)}}/ \mathbb{Z}^{(2)}_2\equiv \mathbb{R}^2 - \{(0,0)\}$ with coordinates $(x,y)$.
Using polar coordinates $x = rcos(\varphi), y = rsin(\varphi)$ the double covering is given by $r = r^+$, $\varphi =2  \varphi^+$ and we obtain the reduced symplectic form $\omega_{\rho^2_{(0,0; y_0)}}=\displaystyle\frac{1}{2}|y_0|^2rdr\wedge d\varphi$.

We have that $\pi_{\rho^2_{(0,0; y_0)}}=D^{2}\circ\pi^{+}_{\rho^2_{(0,0; y_0)}}$ and, besides  that $\pi_{\rho^2_{(0,0; y_0)}}$ is a nontrivial principal bundle with group $\mathbb{Z}^{(2)}\circledS \mathcal{T}_{y_0}$ which is a normal subgroup of $G_{\rho^2_{(0,0; y_0)}}$.

\boldparagraph{\emph{Case} 3.} Let $(q_0,p_0)\in M_{G^2(y_0)}$ with $J(q_0,p_0)=(0,\delta_0y_0)$, $\delta_0\in \mathbb{R}, \delta_0\neq 0$, $y_0\in\mathbb{R}^3-\{0\}$ and $\rho^3_{(0,\delta_0 y_0; y_0)}:=\mathcal{J}(q_0,p_0)$, then
\begin{equation}\label{Caso 3}\begin{array}{rcl}
\mathcal{J}^{-1}(\rho^3_{(0,\delta_0 y_0; y_0)})&=&J^{-1}(0,\delta_0y_0)\cap M_{G^2(y_0)}\vspace{.2cm}\\
&=&\{(\lambda y_0,\beta y_0,\gamma y_0,(\delta_0-\gamma)y_0):\lambda,\beta,\gamma\in\mathbb{R}\}\simeq\mathbb{R}^3,
\end{array}\end{equation}

which is connected so Corollary \ref{corolario formula}  can be applied and the first equality of (\ref{Caso 3})
holds true. This gives a parametrization of $\mathcal{J}^{-1}(\rho^3_{(0,\delta_0 y_0; y_0)})$ and the presymplectic form is $|y_0|^2 d\delta \wedge d\gamma$ where $\delta = \lambda - \beta$.

$G_{\rho^3_{(0,\delta_0 y_0; y_0)}}=G^2(y_0) \circledS \mathcal{T}_{y_0}$.

 $M_{\rho^3_{(0,\delta_0 y_0; y_0)}}\equiv
\mathbb{R}\times\mathbb{R}$. 

The group $G_{\rho^3_{(0,\delta_0 y_0; y_0)}}$ contains $\mathcal{T}_{y_0}$ as a normal subgroup whose action is free and  $G^2(y_0)$ as a subgroup whose action is trivial. These actions commute and $\pi_{\rho^3_{(0,\delta_0 y_0; y_0)}}: \mathcal{J}^{-1}(\rho^3_{(0,\delta_0 y_0; y_0)}) \rightarrow M_{\rho^3_{(0,\delta_0 y_0; y_0)}}$ is a trivial principal bundle isomorphic to $\mathcal{T}_{y_0}\times M_{\rho^3_{(0,\delta_0 y_0; y_0)}}$. A section of this principal bundle giving this trivialization is 
\begin{equation}\label{sigma3}
\sigma_{\rho^3_{(0,\delta_0 y_0; y_0)}} (\delta, \gamma)=(\delta y_0, 0 y_0, \gamma y_0,(\delta_0-\gamma) y_0).
\end{equation}
Using the parametrization the projection $\pi_{\rho^3_{(0,\delta_0 y_0; y_0)}}$ becomes the map $(\lambda, \beta, \gamma)\rightarrow  (\delta, \gamma)$. The reduced symplectic form is $\omega_{\rho^3_{(0,\delta_0 y_0; y_0)}}=|y_0|^2d\delta\wedge d\gamma$.

\boldparagraph{Case 4.} Let $(q_0,p_0)\in M_{G^3(x_0,y_0)}$ with $J(q_0,p_0)=(0,0)$, $x_0, y_0\in\mathbb{R}^3$, $x_0\times y_0\neq 0$ and $\rho^4_{(0,0;x_0,y_0)}:=\mathcal{J}(q_0,p_0)$, then
\begin{equation}\label{Caso 4}
\begin{array}{rcl}
\mathcal{J}^{-1}(\rho^4_{(0,0;x_0,y_0)})&=&J^{-1}(0,0)\cap M_{G^3(x_0,y_0)}\vspace{.2cm}\\
&=&\{(\lambda y_0+x_0,\beta y_0+x_0,0,0): \lambda,\beta\in\mathbb{R}\,\ \lambda\neq\beta\}\vspace{.2cm}\\
&\cup& \{(\lambda y_0+x_0,\beta y_0+x_0,\gamma y_0,-\gamma
y_0):\lambda,\beta,\gamma\in\mathbb{R}, \gamma\neq 0\}\vspace{.2cm}\\
&\simeq&\mathbb{R}^3-\{\ell\},
\end{array}\end{equation}
where $\ell$ is the straight line given by $\gamma=0$ and $\lambda-\beta=0$. which is connected so Corollary \ref{corolario formula}  can be applied and the first equality of (\ref{Caso 4}) holds true. This gives a parametrization of $\mathcal{J}^{-1}(\rho^4_{(0,0;x_0,y_0)})$ in terms of the parameters 
$(\lambda, \delta,\gamma)$, where $\delta = \lambda - \beta$, in the domain $\gamma \neq 0$ or $\delta \neq 0$. The presymplectic form on $\mathcal{J}^{-1}(\rho^4_ {(0,0;x_0, y_0)})$ becomes $|y_0|^2 d\delta \wedge d\gamma$.

 $ G_{\rho^4_{(0,0;x_0,y_0)}}=\{(A,a)\in SE(3):
Ay_0=\pm y_0\ \mbox{and}\ a=\eta y_0+x_0-Ax_0, \eta\in\mathbb{R}\}$.

The subgroup $G^+_{\rho^4_{(0,0;x_0,y_0)}}= G^3(x_0, y_0) \circledS \mathcal{T}_{y_0} $ of $G_{\rho^4_ {(0,0;x_0, y_0)}}$ is a normal subgroup.

If $B \in SO(3)$ satisfies $B^2=I$ and $By_0=-y_0$ then $G^+_{\rho^4_ {(0,0;x_0,y_0)}} \cup \{(B, x_0 – B x_0)\}$
generates $G_{\rho^4_ {(0,0;x_0, y_0)}}$.
Moreover, since the group $\mathbb{Z}^{(4)}_2:=\{(I,0),(B, x_0 – B x_0)\}$ acts by conjugation on $G^+_{\rho^4_ {(0,0;x_0, y_0)}}$, the group $G_{\rho^4_ {(0,0;x_0, y_0)}}$ is naturally isomorphic to the semidirect product $\mathbb{Z}^{(4)}_2\circledS G^+_{\rho^4_{(0,0;x_0, y_0)}}$.

The map $\pi_{\rho^4_ {(0,0;x_0, y_0)}}$ can be obtained in two stages. Using the parametrization, in the first stage we have the bundle $\pi^+_{\rho^4_ {(0,0;x_0, y_0)}}: \mathcal{J}^{-1} (\rho^4_ {(0,0;x_0, y_0)})
\rightarrow M^+_{\rho^4_{(0,0;x_0, y_0)}}$, where $M^+_{\rho^4_ {(0,0;x_0, y_0)}}:=\mathcal{J}^{-1} (\rho^4_ {(0,0;x_0, y_0)})  /G^+_{\rho^4_{(0,0;x_0,y_0)}}\simeq\mathbb{R}^2 - \{(0,0)\}$, with coordinates
$(x^+,y^+) = (\delta, \gamma)$,  or we can use polar coordinates $x^+ = r^+ cos(\varphi^+), y^+ = r^+ sin(\varphi^+)$.

The group $G^{+}_{\rho^4_{(0,0;x_0,y_0)}}$ contains $\mathcal{T}_{y_0}$ as a normal subgroup whose action is free and  $G^3(x_0, y_0)$ as a subgroup whose action is trivial. These actions commute and $\pi^{+}_{\rho^4_{(0,0;x_0, y_0)}}$
is a trivial principal bundle isomorphic to $\mathcal{T}_{y_0}\times M^{+}_{\rho^4_{(0,0;x_0, y_0)}}$. A section $\sigma^+_{\rho^4_ {(0,0;x_0, y_0)}}$ of
$\pi^+_{\rho^4_ {(0,0;x_0, y_0)}}$ giving this trivialization is 
\begin{equation}\label{sigma4}
\sigma^+_{\rho^4_ {(0,0;x_0, y_0)}}(x^+, y^+) = 
(x^+ y_0+x_0, 0 y_0+x_0, y^+ y_0, -y^+ y_0).
\end{equation}
In the second stage, since the action of the group $\mathbb{Z}^{(4)}_2$ on $\mathcal{J}^{-1} (\rho^4_ {(0,0;x_0, y_0)})$ passes to the quotient $M^+_{\rho^4_ {(0,0;x_0, y_0)}}$
we obtain a double covering $D^4:M^+_{\rho^4_ {(0,0;x_0, y_0)}} \longrightarrow M^+_{\rho^4_ {(0,0;x_0, y_0)}}/ \mathbb{Z}^{(4)}_ 2$. Then we obtain $M_{\rho^4_ {(0,0;x_0, y_0)}}=M^+_{\rho^4_ {(0,0;x_0, y_0)}}/ \mathbb{Z}^{(4)}_ 2
\equiv \mathbb{R}^2 - \{(0,0)\}$ with coordinates $(x,y)$.
Using polar coordinates $x = rcos(\varphi)$, $y = rsin(\varphi)$ the double covering is given by $r = r^+, \varphi =2  \varphi^+$ we obtain the reduced symplectic form
$\omega_{\rho^4_{(0,0;x_0,y_0)}}=\displaystyle\frac{1}{2}|y_0|^2rdr\wedge d\varphi$, where $(r,\varphi)$ are the polar coordinates in $M_{\rho^4_{(0,0;x_0,y_0)}}$. 

We have that $\pi_{\rho^4_{(0,0;x_0,y_0)}} = D^{4} \circ \pi^{+}_{\rho^4_{(0,0;x_0,y_0)}}$ and, besides  that $\pi_{\rho^4_{(0,0;x_0,y_0)}}$ is a nontrivial principal bundle with group $\mathbb{Z}^{(2)}\circledS \mathcal{T}_{y_0}$, which is a normal subgroup of $G_{\rho^4_{(0,0;x_0,y_0)}}$. 

Case 4 is related to Case 2 by $\mathcal{J}^{-1} (\rho^4_ {(0,0;x_0, y_0)})= (I, x_0)\mathcal{J}^{-1}(\rho^2_{(0,0; y_0)})$.

\boldparagraph{Case 5.} Let $(q_0,p_0)\in M_{G^3(x_0,y_0)},\ x_0, y_0\in\mathbb{R}^3, z_0:=x_0\times y_0\neq 0$ such that $J(q_0,p_0)=(\delta_0z_0,\delta_0 y_0)$, $\delta_0\in\mathbb{R}$, $\delta_0\neq0$ and $\rho^5_
{(\delta_0 z_0, \delta_0 y_0;x_0, y_0)}:=\mathcal{J}(q_0,p_0)$, then
\begin{equation}\label{Caso 5}\begin{array}{rcl}
\mathcal{J}^{-1}(\rho^5_
{(\delta_0 z_0, \delta_0 y_0;x_0, y_0)})&=&J^{-1}(\delta_0z_0,\delta_0 y_0)\cap M_{G^3(x_0,y_0)}\vspace{.2cm}\\
&=&\{(\lambda y_0+x_0,\beta y_0+x_0,\gamma y_0,(\delta_0-\gamma)
y_0):\lambda,\beta,\gamma\in\mathbb{R}\}\vspace{.2cm}\\
&\cong&\mathbb{R}^3,
\end{array}
\end{equation}

which is connected  so Corollary \ref{corolario formula}  can be applied and the first equality of (\ref{Caso 5}) holds true. This gives a parametrization of $\mathcal{J}^{-1}(\rho^5_
{(\delta_0 z_0, \delta_0 y_0;x_0, y_0)})$ and the presymplectic form is $|y_0|^2 d\delta \wedge d\gamma$ where $\delta = \lambda - \beta$.

 $G_{\rho^5_
{(\delta_0 z_0, \delta_0 y_0;x_0, y_0)}}= G^3(x_0, y_0) \circledS \mathcal{T}_{y_0}$.

$M_{\rho^5_
{(\delta_0 z_0, \delta_0 y_0;x_0, y_0)}}\simeq \mathbb{R}\times\mathbb{R}$.

The group $G_{\rho^5_
{(\delta_0 z_0, \delta_0 y_0;x_0, y_0)}}$ contains 
$\mathcal{T}_{y_0}$ as a normal subgroup whose action is free and  $G^3(x_0 , y_0)$ as a subgroup whose action is trivial. These actions commute and $\pi_{\rho^5_{(\delta_0 z_0, \delta_0 y_0;x_0, y_0)}}: \mathcal{J}^{-1}_{\rho^5_{(\delta_0 z_0, \delta_0 y_0;x_0, y_0)}} \rightarrow M_{\rho^5_{(\delta_0 z_0, \delta_0 y_0;x_0, y_0)}}$ is a trivial principal bundle isomorphic to
$\mathcal{T}_{y_0}\times M_{\rho^5_{(\delta_0 z_0, \delta_0 y_0;x_0, y_0)}}$.
A section of this principal bundle giving the trivialization is
\begin{equation}\label{sigma5}
\sigma_{\rho^5_{(\delta_0 z_0, \delta_0 y_0;x_0, y_0)}} (\delta, \gamma)=(\delta y_0+x_0, 0 y_0+x_0, \gamma y_0, (\delta_0-\gamma) y_0).
\end{equation}
Using the parametrization the projection $\pi_{\rho^5_
{(\delta_0 z_0, \delta_0 y_0;x_0, y_0)}}$ becomes the map $(\lambda, \beta, \gamma)\rightarrow  (\delta, \gamma)$. The reduced symplectic form is $\omega_{\rho^5_
{(\delta_0 z_0, \delta_0 y_0;x_0, y_0)}}=|y_0|^2d\delta\wedge d\gamma$.

Case 5 is related to Case 3 by 
\begin{equation}\label{relacion caso 5 y 3}\mathcal{J}^{-1}(\rho^5_
{(\delta_0 z_0, \delta_0 y_0;x_0, y_0)}) = (I, x_0)\mathcal{J}^{-1}(\rho^3_{(0,\delta_0 y_0; y_0)}).
\end{equation}

\boldparagraph{Case 6.} Let $(q_0,p_0)\in M_{G^4}$ with $J(q_0,p_0)=(\alpha_0,0),\ \alpha_0\neq0$, $\rho^6_{(\alpha_0,0)}:=\mathcal{J}(q_0,p_0)$, and let $\Pi_{\alpha_0}$ be the plane perpendicular to $\alpha_0$ containing $0 \in \mathbb{R}^3$. For each
$p^1 \in \Pi_{\alpha_0} - \{0\}$ there exists a uniquely determined $v_{p^1} \in \Pi_{\alpha_0}$ such that
$v_{p^1}\times p^1 = \alpha_0$ and we have that

$\begin{array}{rcl}
J^{-1}(\alpha_0, 0)&=&\{(q^2+v_{p^1}+\lambda p^1,q^2,p^1,-p^1):q^2 \in \mathbb{R}^3,\ p^1 \in \Pi_{\alpha_0} - \{0\},\ \lambda \in \mathbb{R}\}\vspace{.2cm}\\
&\simeq&\mathbb{R}^3\times(\Pi_{\alpha_0} -\{0\})\times\mathbb{R},\end{array}$

which is connected. Besides we have that $J^{-1}(\alpha_0,0)\cap M_{G^4}=J^{-1}(\alpha_0, 0)$,  so Corollary \ref{corolario formula} can be applied and we have the following parametrization of
$\mathcal{J}^{-1}(\rho^6_{(\alpha_0, 0)}))= J^{-1}(\alpha_0, 0)$ which is a diffeomorphism,  

$P^6_{(\alpha_0,0)}:\mathbb{R}^3 \times (\Pi_{\alpha_0} -\{0\})\times \mathbb{R}\rightarrow \mathcal{J}^{-1}(\rho^6_{(\alpha_0, 0)})$ where $P^6_{(\alpha_0, 0)} (q^2, p^1, \lambda)=( q^2 + v_{p^1} + \lambda  p^1,  q^2,  p^1, - p^1)$.

 $G_{\rho^6_{(\alpha_0,0)}}= G^2(\alpha_0)\circledS\mathbb{R}^3.$

There exists a diffeomorphism $\bold{p}^6_{(\alpha_0, 0)}: \mathbb{R}^+ \times \mathbb{R}\rightarrow M_{\rho^6_{(\alpha_0,0)}}$ such that $\left(\bold{p}^6_{(\alpha_0, 0)}\right)^{-1} \circ \pi_{\rho^6_{(\alpha_0,0)}} \circ P^6_{(\alpha_0, 0)} (q^2, p^1, \lambda)=(|p^1|, \lambda)$. Besides we have that $\pi_{\rho^6_{(\alpha_0, 0)}}$ is a trivial bundle. In fact let
${\bf f}\in \Pi_{\alpha_0} -\{0\}$, $|\bold{f}| = 1$, be fixed. Since any element of
$M_{\rho^6_{(\alpha_0, 0)}}$ is represented as $\bold{p}^6_{(\alpha_0, 0)}(c, \lambda)$  with $c > 0$, $\lambda \in \mathbb{R}$, then 
\begin{equation}\label{sigma6}
\sigma^{\bf f}_{\rho^6_{(\alpha_0,0)}}(\bold{p}^6_{(\alpha_0, 0)}(c, \lambda)):=(v_{c{\bf f}} + \lambda c{\bf f}, 0, c{\bf f}, -c{\bf f})
\end{equation}
represents a section of $\pi_{\rho^6_{(\alpha_0, 0)}}$. 

The bundle $\pi_{\rho^6_{(\alpha_0, 0)}}$ has the trivialization ${\bf P}^6_{(\alpha_0, 0)}
: G_{\rho^6_{(\alpha_0, 0)}} \times M_{\rho^6_{(\alpha_0, 0)}}
\longrightarrow  \mathcal{J}^{-1}(\rho^6_{(\alpha_0, 0)})$ given by the assignment
$\left((A,\xi u_0)), \bold{p}^6_{(\alpha_0, 0)}
(c, \lambda)\right)\rightarrow (A,\xi u_0)\sigma^{\bf f}_{\rho^6_{(\alpha_0, 0)}}\circ \bold{p}^6_{(\alpha_0, 0)}
(c, \lambda)$.

The pullback by $\sigma^{\bf f}_{\rho^6_{(\alpha_0,0)}}$ of the canonical symplectic form on $M$ gives the reduced symplectic form $\omega_{\rho^6_{(\alpha_0,0)}}=\left((\bold{p}^6_{(\alpha_0, 0)})^{-1}\right)^* cd\lambda\wedge dc$.
\ \\
 
\boldparagraph{Case 7.} We shall use the notations introduced in \ref{Apendice B}. Let $(q_0, p_0) \in M_{G^4}$ with $J(q_0, p_0)=(0,u_0),\ u_0\neq 0$ and $\rho^7_{(0,u_0)}:=\mathcal{J}(q_0,p_0)$, then \begin{equation}\label{optimo caso 7}\begin{array}{rcl}
\mathcal{J}^{-1}(\rho^7_{(0,u_0)})&=&J^{-1}(0,u_0)\cap
M_{G^4}\vspace{.2cm}\\
&=&\{(q^1,q^2,p^1,p^2): q^1=r_{q^1}\beta_s+\mu_{q^1}u_0,\ q^2=(r_{q^1}+r_{q^1q^2})\beta_s+\mu_{q^2}u_0,\vspace{.2cm}\\
&& p^1=r_p\beta_s+\lambda u_0,\ p^2=-r_p\beta_s+(1-\lambda)u_0,\ \mbox{with}\ s\in [0,\pi),\vspace{.2cm}\\ 
&& r_{q^1},r_{q^1q^2},\mu_{q^1},\mu_{q^2},\lambda,r_p\in\mathbb{R},\ r^2_{q^1q^2} + r^2_p > 0\ \mbox{and}\vspace{.2cm}\\
&& r_{q^1}+r_{q^1q^2}(1-\lambda)+r_p(\mu_{q^2}-\mu_{q^1})=0\}.
\end{array}
\end{equation}

By defining $r>0$ and $\tau \in [0, 2\pi)$ by $r_{q^1 q^2} = r\cos(\tau)$, $r_p = r\sin(\tau)$, formula (\ref{optimo caso 7}) 
gives, in particular, an injective map $P^7_{(0,u_0)}: \mathbb{R}P^1 \times  S^1 \times \mathbb{R}^+ \times \mathbb{R}
\times \mathbb{R}\times \mathbb{R} \rightarrow M$,
say $(q,p) = P^7_{(0,u_0)}(\ell_s, e^{i\tau}, r, \mu_{q^1}, \mu_{q^2},\lambda)$. The image of $P^7_{(0,u_0)}$ coincides with the r.h.s of the first equation of (\ref{optimo caso 7}), namely $J^{-1}(0,u_0)\cap M_{G^4}$, and it is a connected subset of $M$. The restrictions of $P^7_{(0,u_0)}$ to
$(\mathbb{R}P^1-  \{\ell_0\}) \times  S^1 \times \mathbb{R}^+ \times \mathbb{R}
\times \mathbb{R}\times \mathbb{R}$ and to $\{\ell_0\} \times  S^1 \times \mathbb{R}^+ \times \mathbb{R}
\times \mathbb{R}\times \mathbb{R}$ are continuous. \textit{Warning:} $P^7_{(0,u_0)}$ is not continuous at points where
$\ell_s = \ell_0$. This follows from the fact that the curve described by the point $(\ell_s, e^{i\tau}, r, \mu_{q^1}, \mu_{q^2}\lambda) \in\mathbb{R}P^1 \times  S^1 \times \mathbb{R}^+ \times \mathbb{R}\times \mathbb{R}\times \mathbb{R}$ where $\tau, r, \mu_{q^1}, \mu_{q^2}$ are fixed and $s\in[0,\pi)$, is smooth and closed while its image  $(q(s),p(s))$ under  $P^7_{(0,u_0)}$ satisfies that $(q(0),p(0))$ is different from the limit of $(q(s),p(s))$  as $s$ tends to $\pi$, in particular, using the notation of \ref{Apendice B}, $r_{f_{\pi}(p(0))} = -r_{p(0)}$ and  $r_{f_{\pi}(q^1(0))f_{\pi}(q^2(0))} = -r_{q^1(0)q^2(0)}$.

The optimal momentum isotropy subgroup is $G_{\rho^7_{(0,u_0)}}=G^2(u_0) \circledS \mathcal{T}_{u_0}$, and it acts freely and properly on $\mathcal{J}^{-1} (\rho^7_{(0, u_0)})$ so, $\pi_{\rho^7_{(0, u_0)}}:
\mathcal{J}^{-1} (\rho^7_{(0, u_0)}) \longrightarrow
M_{\rho^7_{(0, u_0)}}$ is a principal bundle. There exists a diffeomorphism $\bold{p}^7_{(0,u_0)}: \mathcal{P}(\Pi_0) \times\mathbb{R}^+ \times \mathbb{R}\times \mathbb{R} \rightarrow 
M_{\rho^7_{(0,u_0)}}$ such that $(\bold{p}^7_{(0,u_0)})^{-1}\circ \pi_{\rho^7_{(0,u_0)}}\circ P^7_{(0,u_0)}(\ell_s,  e^{i\tau}, r, \mu_{q^1}, \mu_{q^2}, \lambda)= ( \ell_{[\tau]_{\pi}}^{\Pi_0}, r , \mu_{q^2} - \mu_{q^1}, \lambda)$. 

Let ${\bf \sigma}:\mathbb{R}P^1\times \mathbb{R}^+ \times \mathbb{R}\times \mathbb{R}\longrightarrow\mathbb{R}P^1 \times  S^1 \times \mathbb{R}^+ \times \mathbb{R}\times \mathbb{R}\times \mathbb{R}$ given by ${\bf \sigma}(\ell_{s},r,\mu, \lambda)=(\ell_{s}, e^{is}, r, 0, \mu, \lambda)$. Then $({\bf p}^7_{(0,u_0)})^{-1} \circ\pi_{\rho^7_{(0,u_0)}}\circ  P^7_{(0,u_0)} \circ {\bf \sigma}\circ \varphi^{\Pi_0}(\ell^{\Pi_0}_{\nu},  r, \mu, \lambda)=(\ell^{\Pi_0}_{\nu},r,\mu,\lambda)$ and therefore 
\begin{equation}\label{sigma7}
\sigma_{\rho^7_{(0,u_0)}}:=P^7_{(0,u_0)} \circ {\bf \sigma}\circ \varphi^{\Pi_0}\circ ({\bf p}^7_{(0,u_0)})^{-1}
\end{equation}
is a section of $\pi_{\rho^7_{(0,u_0)}}$ . The bundle $\pi_{\rho^7_{(0,u_0)}}$ is a trivial principal bundle where the trivialization ${\bf P}^7_{(0,u_0)}
: G_{\rho^7_{(0,u_0)}} \times M_{\rho^7_{(0,u_0)}}
\longrightarrow  \mathcal{J}^{-1}(\rho^7_{(0,u_0)})$ is given by the assignment
$\left((A,\xi u_0)), \bold{p}^7_{(0,u_0)}
(\ell_{\nu}^{\Pi_0}, r, \mu, \lambda)\right)\rightarrow (A,\xi u_0)\sigma_{\rho^7_{(0,u_0)}}\circ \bold{p}^7_{(0,u_0)}
(\ell_{\nu}^{\Pi_0}, r, \mu, \lambda)$.

Using the diffeomorphism $\bold{p}^7_{(0, u_0)}$, the
reduced symplectic form which is equal to
$\sigma_{\rho^7_{(0,u_0)}}^*\Omega_M$ is represented by $({\bf p}^7_{(0,u_0)})^*\omega_{\rho^7_{(0,u_0)}} = 
-rdr \wedge d\nu-u ^2_{03} d\mu \wedge d\lambda$.

Let $\mathrm{p}:S^1\times S^1\times\mathbb{R}^+\times\mathbb{R}\times\mathbb{R}\times\mathbb{R}\rightarrow\mathbb{R}P^1\times S^1\times\mathbb{R}^+\times\mathbb{R}\times\mathbb{R}\times\mathbb{R}$ given by $\mathrm{p}(e^{i\tilde{s}},e^{i\tau},r,\mu_{q^1},\mu_{q^2},\lambda)=(\ell_{[\tilde{s}]_{\pi}},e^{i\tau},r,\mu_{q^1},\mu_{q^2},\lambda)$. Then $\widetilde{P}^7_{(0,u_0)}:=P^7_{(0,u_0)}\circ\mathrm{p}$ is a double covering of $\mathcal{J}^{-1}(\rho^7_{(0,u_0)})$.

\boldparagraph{Case 8.} We shall use the results of \ref{Apendice B} and \ref{Apendice C}. Let $(q_0,p_0)\in M_{G^4}$ with $J(q_0,p_0)=(\delta_0 u_0,u_0)$,\ $\delta_0\neq 0$, $u_0\neq0$ and $\rho^8_{(\delta_0 u_0, u_0)}:=\mathcal{J}(q_0,p_0)$,  then $\mathcal{J}^{-1}(\rho^8_{(\delta_0 u_0, u_0)})=J^{-1}(\delta_0 u_0,u_0)\cap M_{G^4}=J^{-1}(\delta_0 u_0,u_0)$. 

If $(q,p)\in J^{-1}(\delta_0 u_0,u_0)$, then $N:=p^1\times p^2 \neq 0$ and we denote $q^1_N$ and 
$q^2_N$ the orthogonal projections of $q^1$ and $q^2$ on the plane $\pi_{N}$ perpendicular to $N$. The points $(q_N,p)=(q^1_N, q^2_N, p^1, p^1)$ belong to the chart $W_{\mathfrak{p}} \subseteq \mathcal{J}^{-1} (\rho^7_{(0, u_0)})$ defined in \ref{Apendice C}. In particular the parametrization  $P^7_{(0,u_0)}$ given in (\ref{optimo caso 7}) holds for those points and therefore $N = r_p \beta_s \times u_0$. In other words we have a parametrization $P^8_{(\delta_0u_0, u_0)}:=\mathcal{F}_{\mathfrak{p}} \circ P^7_{(0, u_0)}| _{\mathbb{R}P^1 \times (S^1 -\{e^{i0}, e^{i\pi}\})\times \mathbb{R}^+ \times \mathbb{R}\times \mathbb{R} \times \mathbb{R}}$ of $\mathcal{J}^{-1}(\rho^8_{(\delta_0 u_0, u_0)})$ where the diffeomorphism
$\mathcal{F}_{\mathfrak{p}}: W_{\mathfrak{p}} \rightarrow\mathcal{J}^{-1} (\rho^8_{(\delta_0 u_0, u_0)})$ is given by $\mathcal{F}_{\mathfrak{p}}(q^1_N, q^2_N, p^1, p^2)=(q^1_N + \tau^1 N, q^2_N+ \tau^2 N, p^1, p^2)$ where $\tau^1=\displaystyle\frac{\delta_0}{r_p^2}(1-\lambda)$, $\tau^2=-\displaystyle\frac{\delta_0\lambda}{r_p^2}$ . Besides, the momentum isotropy subgroups $G_{\rho^7_{(0, u_0)}}$  and $G_{\rho^8_{(\delta_0 u_0 , u_0)}}$ coincide and $\mathcal{F}_{\mathfrak{p}}$ is equivariant. There is a diffeomorphism $\mathcal{F}_{\mathfrak{p},\rho^7_{(0,u_0)}}: \pi_{\rho^7_{(0,u_0)}}(W_{\mathfrak{p}})
 \longrightarrow  M_{\rho^8_{(\delta_0 u_0, u_0)}}$ such that
 \begin{equation}\label{formula Fp}
 \pi_{\rho^8_{(\delta_0 u_0 , u_0)}}\circ \mathcal{F}_{\mathfrak{p}}=\mathcal{F}_{\mathfrak{p},\rho^7_{(0,u_0)}}\circ \pi_{\rho^7_{(0 , u_0)}}\vert_{W_{\mathfrak{p}}}.
 \end{equation}
 
Define ${\bf p}^8 _{(\delta_0 u_0, u_0)}: (\mathcal{P}(\Pi_0) -\{\ell_0^{\Pi_0}\})\times \mathbb{R}^{+}\times \mathbb{R}\times \mathbb{R}
\longrightarrow M_{\rho^8_{(\delta_0 u_0, u_0)}}$ as follows ${\bf p}^8 _{(\delta_0 u_0, u_0)} :=\mathcal{F}_{\mathfrak{p}, \rho^7_{(0, u_0)}}\circ {\bf p}^7 _{(0, u_0)}|_{(\mathcal{P} (\Pi_0) - \{\ell_0^{\Pi_0}\})\times \mathbb{R}^{+}\times \mathbb{R}\times \mathbb{R}}$.
 
The map 
\begin{equation}\label{sigma8}
{\rho^8_{(\delta_0 u_0,u_0)}}:=\mathcal{F}_{\mathfrak{p}} \circ \sigma_{\rho^7_{(0, u_0)}} \circ [\mathcal{F}_{\mathfrak{p}, \rho^7_{(0, u_0)}}]^{-1}
\end{equation}
from $M_{\rho^8_{(\delta_0 u_0, u_0)}}$ to $\mathcal{J}^{-1} (\rho^8_{(\delta_0 u_0, u_0)})$ is a section of 
$\pi_{\rho^8_{(\delta_0 u_0,u_0)}}$. The following equalities hold 
\vspace{.2cm}

$\left(\sigma_{\rho^8_{(\delta_0 u_0,u_0)}}\circ \mathcal{F}_{\mathfrak{p}, \rho^7_{(0, u_0)}}\circ (\bold{p}^7_{(0,u_0)}\vert_{\mathcal{P}(\Pi_0)-\{\ell_0^{\Pi_0}\}})\right)^*\Omega_M$
\begin{equation}\begin{array}{rcl}\label{omega 8}
&=&\left(\mathcal{F}_{\mathfrak{p}}\circ\sigma_{\rho^7_{(0, u_0)}}\circ (\bold{p}^7_{(0,u_0)}\vert_{\mathcal{P}(\Pi_0)-\{\ell_0^{\Pi_0}\}})\right)^*\Omega_M\vspace{.2cm}\\
&=&\left(\sigma_{\rho^7_{(0,u_0)}}\circ (\bold{p}^7_{(0,u_0)}\vert_{\mathcal{P}(\Pi_0)-\{\ell_0^{\Pi_0}\}})\right)^*\Omega_M\vspace{.2cm}\\
&=&\displaystyle-\displaystyle r_N dr_N\wedge d\nu_N-u^2_{03}d\mu_N\wedge d\lambda,
\end{array}\end{equation}

where $\nu_N \in [0,\pi)$, $r^2_N=r^2_p+r^2 _{q^1_N q^2_N}$ and $\mu_N = \mu_{q^2_N}-\mu_{q^1_N}$.

The map $\mathcal{F}_{\mathfrak{p},\rho^7_{(0,u_0)}}$ is a symplectomorphism and $\mathcal{F}_{\mathfrak{p}}$ is a presymplectomorphism where the presymplectic forms $\omega_{W_{\mathfrak{p}}}$ and $\omega_{\mathcal{J}^{-1}(\rho^8_{(\delta_0 u_0, u_0)})}$ are the pullback of $\Omega_M$ via the embeddings of $W_{\mathfrak{p}}$ and $\mathcal{J}^{-1}(\rho^8_{(\delta_0 u_0, u_0)})$ in $M$, respectively. We have $\left( {\bf p}^8_{(\delta_0 u_0, u_0)}\right)^*\omega_{\rho^8_{(\delta_0 u_0, u_0)}}=-r_{N}dr_{N}\wedge d\nu_{N} – u^2_{03} d\mu_{N} \wedge d\lambda$.

The bundle $\pi_{\rho^8_{(\delta_0 u_0,u_0)}}$ is a trivial principal bundle where the trivialization ${\bf P}^8_{(\delta_0 u_0,u_0)}
: G_{\rho^8_{(\delta_0 u_0,u_0)}} \times M_{\rho^8_{(\delta_0 u_0,u_0)}}
\longrightarrow  \mathcal{J}^{-1}(\rho^8_{(\delta_0 u_0,u_0)})$ is given by the assignment
$\left((A,\xi u_0)), \bold{p}^8_{(\delta_0 u_0,u_0)}
(\ell_{\nu_N}^{\Pi_0}, r_N, \mu_N, \lambda)\right)\rightarrow (A,\xi u_0)(\sigma_{\rho^8_{(\delta_0 u_0,u_0)}}\circ {\bf p}^8_{(\delta_0 u_0,u_0)})
(\ell_{\nu_N}^{\Pi_0}, r_N, \mu_N, \lambda)$.

Using the diffeomorphism $\bold{p}^8_{(\delta_0 u_0, u_0)}$, the
reduced symplectic form which is equal to
$\sigma_{\rho^8_{(\delta_0 u_0,u_0)}}^*\Omega_M$ is represented by $({\bf p}^8_{(\delta_0 u_0,u_0)})^*\omega_{\rho^8_{(\delta_0 u_0,u_0)}} = 
-r_Ndr_N \wedge d\nu_N-u ^2_{03} d\mu_N \wedge d\lambda$.
\ \\

Before we state Case 9 and Case 10 we will introduce some notation.
For $a \in \mathbb{R}^3$ let ${\bf T}_{a} :  M \longrightarrow M$
be the symplectomorphism given by ${\bf T}_a(q,p)=(I,a)(q,p)$, and, also, ${\bf C}_a:SO(3)\circledS \mathbb{R}^3 \longrightarrow SO(3) \circledS \mathbb{R}^3$  the group isomorphism given by conjugation ${\bf C}_a (B,b) = (I,a) (B,b) (I,-a) = (B, b + a - Ba)$.

The following equality holds ${\bf T}_a \left((B,b)(q,p)\right) 
= {\bf C}_a (B,b) {\bf T}_a (q,p)$



\boldparagraph{Case 9.} Let $(q_0,p_0)\in M_{G^4}$ with $J(q_0,p_0)=(\alpha_0,u_0)$, $\alpha_0\times u_0\neq 0$, $<\alpha_0,u_0>=0$, $\rho^9_{(\alpha_0, u_0)}:=\mathcal{J}(q_0,p_0)$, and let $a^9_{(\alpha_0, u_0)}$ be the vector  uniquely determined by the conditions $<a^9_{(\alpha_0, u_0)},u_0>=0$ and $a^9_{(\alpha_0, u_0)}\times u_0=\alpha_0$. Then $\mathcal{J}^{-1}(\rho^9_{(\alpha_0, u_0)})=
{\bf T}_{a^9_{(\alpha_0, u_0)}}\left( \mathcal{J}^{-1}(\rho^7_{(0,u_0)})\right)$, and therefore, using Case 7, we have the parametrization of $\mathcal{J}^{-1}(\rho^9_{(\alpha_0, u_0)})$ given by 
$P^9_{(\alpha_0, u_0)}:\mathbb{R}P^1 \times  S^1 \times  \mathbb{R}^+ \times  \mathbb{R}\times \mathbb{R}\times \mathbb{R}\rightarrow M$, where $P^9_{(\alpha_0, u_0)} = \bold{T}_{a^9_{(\alpha_0, u_0)}} \circ P^7_{(0, u_0)}$.

The map, ${\bf T}_{a^9_{(\alpha_0, u_0)}} |_{\mathcal{J}^{-1}(\rho^7_{(0,u_0)})}$ is a diffeomorphism and also a presymplectomorphism onto $\mathcal{J}^{-1}(\rho^9_{(\alpha_0, u_0)})$. The momentum isotropy subgroup $G_{\rho^9_{(\alpha_0, u_0)}}$ satisfies that ${\bf C}_{a^9_{(\alpha_0, u_0)}}|_{G_{\rho^7_{(0,u_0)})}}:G_{\rho^7_{(0,u_0)})}
\rightarrow G_{\rho^9_{(\alpha_0, u_0)}}$ is a group isomorphism. The condition $\mathcal{J}^{-1}(\rho^9_{(\alpha_0, u_0)})=J^{-1}(\alpha_0, u_0)\cap M_{G^4}$ is satisfied.

There exists a symplectomorphism ${\bf t}_{a^9_{(\alpha_0, u_0)}}: M_{\rho^7_{(0,u_0)})}\rightarrow M_{\rho^9_{(\alpha_0, u_0)}}$
such that $\pi_{\rho^9_{(\alpha_0, u_0)}}\circ {\bf T}_{a^9_{(\alpha_0, u_0)}}|_{\mathcal{J}^{-1}(\rho^7_{(0,u_0)})} =
{\bf t}_{a^9_{(\alpha_0, u_0)}} \circ \pi_{\rho^7_{(0,u_0)}}$. Define $\bold{p}^9 _{(\alpha_0, u_0)}: \mathcal{P}(\Pi_0) \times \mathbb{R}^{+}\times \mathbb{R}\times \mathbb{R}\longrightarrow M_{\rho^9_{(\alpha_0, u_0)}}$
as follows $\bold{p}^9 _{(\alpha_0, u_0)} :=\bold{t}_{a^9_{(\alpha_0, u_0)}}\circ \bold{p}^7 _{(0, u_0)}$.

A section $\sigma_{\rho^9_{(\alpha_0, u_0)}} : M_{\rho^9_{(\alpha_0, u_0)}} \rightarrow  \mathcal{J}^{-1}(\rho^9_{(\alpha_0, u_0)})$ is defined as follows
\begin{equation}\label{sigma9}
\sigma_{\rho^9_{(\alpha_0, u_0)}} = {\bf T}_{a^9_{(\alpha_0, u_0)}}\circ \sigma_{\rho^7_{(0, u_0)}}\circ
\left({\bf t}_{a^9_{(\alpha_0, u_0)}}\right)^{-1}.\end{equation}
We can represent $\omega_{\rho^9_{(\alpha_0, u_0)}}$ as a $2$-form on
$\mathbb{R}P^1 \times \mathbb{R}^+\times \mathbb{R} \times\mathbb{R}$ namely $\left(\sigma_{\rho^9_{(\alpha_0, u_0)}}\circ \bold{p}^9 _{(\alpha_0, u_0)}\right)^* \Omega_M$ which coincides with
$-rdr \wedge d\nu-u ^2_{03} d\mu \wedge d\lambda$.

The bundle $\pi_{\rho^9_{(\alpha_0, u_0)}}$ is a trivial principal bundle where the trivialization ${\bf P}^9_{(\alpha_0, u_0)}
: G_{\rho^9_{(\alpha_0, u_0)}} \times M_{\rho^9_{(\alpha_0, u_0)}}
\longrightarrow  \mathcal{J}^{-1}(\rho^9_{(\alpha_0, u_0)})$ is given by the assignment
$\left(\bold{C}_{a^9_{(\alpha_0, u_0)}} (A, \xi u_0), \bold{p}^9_{(\alpha_0, u_0)}
(\ell_{\nu}^{\Pi_0}, r, \mu, \lambda)\right)\rightarrow \bold{C}_{a^9_{(\alpha_0, u_0)}} (A, \xi u_0)(\sigma_{\rho^9_{(\alpha_0, u_0)}}\circ {\bf p}^9_{(\alpha_0, u_0)})
(\ell_{\nu}^{\Pi_0}, r, \mu, \lambda)$.

We have $({\bf p}^9_{(\alpha_0, u_0)})^*\omega_{\rho^9_{(\alpha_0, u_0)}} = 
-rdr \wedge d\nu-u ^2_{03} d\mu \wedge d\lambda$.

\boldparagraph{Case 10.} Let $(q_0,p_0)\in M_{G^4}$ with $J(q_0,p_0)=(\alpha_0,u_0)$,\ $\alpha_0\times u_0\neq 0,
<\alpha_0,u_0> \neq 0$, $\rho^{10}_{(\alpha_0, u_0)}:=\mathcal{J}(q_0,p_0)$, and let $a^{10}_{(\alpha_0, u_0)}$ be the vector and $\delta^{10}_{(\alpha_0, u_0)}\neq 0$ the number uniquely determined by the conditions $<a^{10}_{(\alpha_0, u_0)},u_0>=0$ and $a^{10}_{(\alpha_0, u_0)}\times u_0+\delta^{10}_{(\alpha_0, u_0)} u_0= \alpha_0$. Then $\mathcal{J}^{-1}(\rho^{10}_{(\alpha_0, u_0)})={\bf T}_{a^{10}_{(\alpha_0,u_0)}}\left(\mathcal{J}^{-1}(\rho^8_{(\delta^{10}_{(\alpha_0, u_0)} u_0, u_0)})\right)$, and therefore, using Case 8, we have the parametrization of $\mathcal{J}^{-1}(\rho^{10}_{(\alpha_0, u_0)})$ given by $P^{10}_{(\alpha_0, u_0)} = {\bf T}_{a^{10}_{(\alpha_0, u_0)}}\circ 
P^8_{(\delta^{10}_{(\alpha_0, u_0)} u_0, u_0)}$.

The map, ${\bf T}_{a^{10}_{(\alpha_0, u_0)}} |_{\mathcal{J}^{-1}(\rho^8_{(\delta^{10}_{(\alpha_0, u_0)} u_0, u_0)})}$ is a diffeomorphism and also a presymplectomorphism onto $\mathcal{J}^{-1}(\rho^{10}_{(\alpha_0, u_0)})$. The momentum isotropy subgroup $G_{\rho^{10}_{(\alpha_0, u_0)}}$ satisfies that ${\bf C}_{a^{10}_{(\alpha_0, u_0)}}|_{G_{\rho^8_{(\delta^{10}_{(\alpha_0, u_0)} u_0, u_0)})}}:G_{\rho^8_{(\delta^{10}_{(\alpha_0, u_0)} u_0, u_0)})}
\rightarrow G_{\rho^{10}_{(\alpha_0, u_0)}}$ is a group isomorphism. The condition $\mathcal{J}^{-1}(\rho^{10}_{(\alpha_0, u_0)})=J^{-1}(\alpha_0, u_0)\cap M_{G^4}$ is satisfied, which comes from the similar formula in Case 8.

There exists a symplectomorphism ${\bf t}_{a^{10}_{(\alpha_0, u_0)}}: M_{\rho^8_{(\delta^{10}_{(\alpha_0, u_0)} u_0, u_0)})}\rightarrow M_{\rho^{10}_{(\alpha_0, u_0)}}$
such that $\pi_{\rho^{10}_{(\alpha_0, u_0)}}\circ {\bf T}_{a^{10}_{(\alpha_0, u_0)}}|_{\mathcal{J}^{-1}(\rho^8_{(\delta^{10}_{(\alpha_0, u_0)} u_0, u_0)})} =
{\bf t}_{a^{10}_{(\alpha_0, u_0)}} \circ \pi_{\rho^8_{(\delta^{10}_{(\alpha_0, u_0)} u_0, u_0)}}$. Define $\bold{p}^{10} _{(\alpha_0, u_0)}: (\mathcal{P}(\Pi_0)- \{\ell_0^{\Pi_0}\}) \times \mathbb{R}^{+}\times \mathbb{R}\times \mathbb{R}\longrightarrow M_{\rho^{10}_{(\alpha_0, u_0)}}$
as follows $\bold{p}^{10} _{(\alpha_0, u_0)} :=\bold{t}_{a^{10}_{(\alpha_0, u_0)}}\circ \bold{p}^8 _{(\delta^{10}_{(\alpha_0, u_0)} u_0, u_0)}$.

A section $\sigma_{\rho^{10}_{(\alpha_0, u_0)}} : M_{\rho^{10}_{(\alpha_0, u_0)}} \rightarrow  \mathcal{J}^{-1}(\rho^{10}_{(\alpha_0, u_0)})$ is defined as follows
\begin{equation}
\sigma_{\rho^{10}_{(\alpha_0, u_0)}} = {\bf T}_{a^{10}_{(\alpha_0, u_0)}}\circ \sigma_{\rho^8_{(\delta^{10}_{(\alpha_0, u_0)} u_0, u_0)}}\circ
\left({\bf t}_{a^{10}_{(\alpha_0, u_0)}}\right)^{-1}.
\end{equation}
The reduced symplectic form is $\omega_{\rho^{10}_{(\alpha_0, u_0)}} = 
\sigma_{\rho^{10}_{(\alpha_0, u_0)}}^* \Omega_M$. We can represent $\omega_{\rho^{10}_{(\alpha_0, u_0)}}$ as a $2$-form on
$(\mathcal{P}(\Pi_0)-\{\ell_0^{\Pi_0}\}) \times \mathbb{R}^+\times \mathbb{R} \times\mathbb{R}$ namely $\left(\sigma_{\rho^{10}_{(\alpha_0, u_0)}}\circ {\bf p}^{10}_{(\alpha_0, u_0)}\right)^* \Omega_M$ which coincides with
$-r_Ndr_N \wedge d\nu_N-u ^2_{03} d\mu_N \wedge d\lambda$.

The bundle $\pi_{\rho^{10}_{(\alpha_0, u_0)}}$ is a trivial principal bundle where the trivialization ${\bf P}^{10}_{(\alpha_0, u_0)}
: G_{\rho^{10}_{(\alpha_0, u_0)}} \times M_{\rho^{10}_{(\alpha_0, u_0)}}
\longrightarrow  \mathcal{J}^{-1}(\rho^{10}_{(\alpha_0, u_0)})$ is given by the assignment
$\left(\bold{C}_{a^{10}_{(\alpha_0, u_0)}} (A, \xi u_0), \bold{p}^{10}_{(\alpha_0, u_0)}
(\ell_{\nu}^{\Pi_0}, r, \mu, \lambda)\right)\rightarrow \bold{C}_{a^{10}_{(\alpha_0, u_0)}} (A, \xi u_0)(\sigma_{\rho^{10}_{(\alpha_0, u_0)}}\circ {\bf p}^{10}_{(\alpha_0, u_0)})
(\ell_{\nu}^{\Pi_0}, r, \mu, \lambda)$.
\ \\

\item For each $(\alpha_0,u_0)\in \mathbb{R}^3\times\mathbb{R}^3$ and $(q,p)\in J^{-1}(\alpha_0,u_0)$, the condition that if $J^{-1}(\alpha_0,u_0)\cap M_{SE(3)_{(q,p)}}$ is non empty then it is connected, is satisfied and then from Corollary \ref{corolario formula}  it follows that $\mathcal{J}{(q,p)}=J^{-1}(\alpha_0,u_0)\cap M_{SE(3)_{(q,p)}}$. In particular, the notations for the values of the optimal momentum map introduced in (A) and in Corollary \ref{corolario formula} are compatible in the following way:
$\rho^1_{(0,0;x_0)}=\rho_{((0,0),G^1(x_0))}$; $\rho^2_{(0,0;y_0)}=\rho_{((0,0),G^2(y_0))}$; $\rho^3_{(0,\delta_0y_0;y_0)}=\rho_{((0,\delta_0 y_0),G^2(y_0))}$; $\rho^4_{(0,0;x_0,y_0)}=\rho_{((0,0),G^3(x_0, y_0))}$; $\rho^5_{(\delta_0(x_0\times y_0), \delta_0y_0;x_0,y_0)}=\rho_{((\delta_0(x_0\times y_0),\delta_0 y_0),G^3(x_0, y_0))}$; $\rho^6_{(\alpha_0,0)}=\rho_{((\alpha_0,0),G^4)}$; $\rho^7_{(0,u_0)}=\rho_{((0, u_0),G^4)}$; $\rho^8_{(\delta_0 u_0, u_0)}=\rho_{((\delta_0 u_0, u_0),G^4)}$; $\rho^9_{(\alpha_0, u_0)}=\rho_{((\alpha_0, u_0), G^3(a, u_0))}$ with $\alpha_0 \times u_0 \neq 0$, $<\alpha_0, u_0> = 0$; $\rho^{10}_{(\alpha_0, u_0)}:=\rho_{((\alpha_0, u_0), G^3(a, u_0))}$ with $\alpha_0 \times u_0 \neq 0$, $<\alpha_0, u_0> \neq 0$. 

The subscripts of the previous expressions give a parametrization of the values of the optimal momentum which is not injective in Cases 2,3,4,5. An injective parametrization for those cases is obtained using the notation given in the paragraph Parametrization of classes of isotropy subgroups as follows
$\rho^2_{(0,0;[y_0]});\ \rho^3_{(0,\delta_0 y_0 ;[y_0])};\ 
\rho^4_{(0,0;[x_0 , y_0])};\ 
\rho^5_{(\delta_0(x_0\times y_0), \delta_0y_0;[x_0,y_0])}$. We will use both notations.
\end{enumerate}
\end{theorem}

\boldparagraph{\emph{Proof of (A)}}

\ \\\emph{Proof of Cases 1, 2, 3, 4 and 5.}

The proof of the statements of \emph {Case 1}, \emph{Case 2} and \emph{Case 3} can be done in a direct way. The level set of the optimal momentum map for \emph{Case 4} and \emph{Case 5} can be obtained using left translation by $(I,x_0)$ of the level set of the optimal momentum map for \emph{Case 2} and \emph{Case 3} respectively, which can be proven using the facts that $J^{-1}(0,0)=(I,x_0)J^{-1}(0,0)$; $J^{-1}(\delta_0 (x_0\times y_0),\delta_0 y_0)=(I, x_0) J^{-1}(0,\delta_0 y_0)$; $M_{G^3(x_0, y_0)}=(I, x_0) M_{G^2(y_0)}$. Therefore  the validity of the  statements for \emph{Case 4} and \emph{Case 5} can be obtained, after some work, using formula (\ref{equivariance optimal}), from the validity of the statements of \emph{Case 2} and \emph{Case 3}, respectively.

\ \\\emph{Proof of Case 6.}

From (\ref{preimagen del momento alpha}), (\ref{preimagen del momento u}) we obtain in this case $(q^1-q^2)\times p^1=\alpha_0$ and from this and the fact that $\alpha_0\neq 0$ it follows that all the solutions of (\ref{preimagen del momento alpha}) and (\ref{preimagen del momento u}) belong to $M_3$ and then also to $M_{G^4}$.
Therefore
$J^{-1}(\alpha_0,0)\cap M_{G^4}=J^{-1}(\alpha_0,0)$. So in this case, because of Corollary \ref{corolario formula}, the level set of the optimal momentum map and the level set of the momentum map coincide provided that the latter is connected, which, in turn, follows from the parametrization. 
The group $G_{\rho^6_{(\alpha_0, 0)}}$ acts freely on $\mathcal{J}^{-1}(\rho^6_{(\alpha_0, 0)}))$. Besides, $|p^1|$ and $\lambda$ remain constant on each orbit of this action and, moreover, each pair $(|p^1|, \lambda)$ determines biunivocally the orbit. We can deduce that $M_{\rho^6_{(\alpha_0, 0)}}$ is diffeomorphic to $\mathbb{R}^+ \times \mathbb{R}$.
The rest of the proof is not difficult and will be omitted.

\ \\\emph{Proof of Case 7.}

Injectiveness of $P^7_{(0,u_0)}$ and the fact that 
the restrictions of $P^7_{(0,u_0)}$ to $(\mathbb{R}P^1-\{\ell_0\}) \times  S^1 \times \mathbb{R}^+ \times \mathbb{R}
\times \mathbb{R}\times \mathbb{R}$ and to
$\{\ell_0\} \times  S^1 \times \mathbb{R}^+ \times \mathbb{R}
\times \mathbb{R}\times \mathbb{R}$ are continuous can be proven using the definitions.  The statement about the limit as $s$ tends to $\pi$  of the curve $(\ell_s, e^{i\tau}, r, \mu_{q^1}, \mu_{q^2}\lambda)$ was proven essentially in \ref{Apendice B}.

In order to prove the second equality of  formula (\ref{optimo caso 7}) one first shows, using basic Euclidean geometry in $\mathbb{R}^3$, that if $q^1, q^2,p^1, p^2$ satisfy 
 formulas (\ref{preimagen del momento alpha}) and (\ref{preimagen del momento u}) with angular momentum $\alpha=0$ and linear momentum $u$ not $0$ then $q^1$, $q^2$, $p^1$, $p^2$ belong to some $\Pi_s$, $s\in [0,\pi)$. This leads to proving directly the existence of parameters $r_{q^1q^2}$, $r_p$, $r_{q^1}$, $\mu_{q^1}$, $\mu_{q^2}$, satisfying the required equations appearing in the r.h.s of the second equality of (\ref{optimo caso 7}). The inequations are proven using directly that $(q, p) \in M_{G^4}$. Conversely, one can show directly that any point $(q,p) \in M$ satisfying the equations and inequations that appear in the r.h.s of the second equation of (\ref{optimo caso 7}), for some values of the parameters, must belong to the r.h.s of the first equation of (\ref{optimo caso 7}). We have proven that the image of $P^7_{(0, u_0)}$ coincides with $J^{-1}(0, u_0)\cap M_{G^4}$. 
 
Now we shall prove that $J^{-1}(0, u_0)\cap M_{G^4}$ is arc-connected. First, we can see using the definitions and what has been proven so far, that $J^{-1}(0, u_0)\cap M_{G^4} \cap  \Pi_0 = 
P^7_{(0,u_0)}(\{\ell_0\} \times  S^1 \times \mathbb{R}^+ \times \mathbb{R}\times \mathbb{R}\times \mathbb{R})$. Since 
$P^7_{(0,u_0)}$ restricted to $\{\ell_0\} \times  S^1 \times \mathbb{R}^+ \times \mathbb{R}\times \mathbb{R}\times \mathbb{R}$ is continuous and $\{\ell_0\} \times  S^1 \times \mathbb{R}^+ \times \mathbb{R}\times \mathbb{R}\times \mathbb{R}$ is arc-connected we can conclude that $J^{-1}(0, u_0)\cap M_{G^4} \cap  \Pi_0$ is  arc-connected. Now let $(q,p) \in
J^{-1}(0, u_0)\cap M_{G^4}\cap \Pi_s, s \in [0,\pi)$, say
$(q,p) = P^7_{(0,u_0)}(\ell_s, e^{i\tau}, r, \mu_{q^1}, \mu_{q^2},\lambda)$. A rotation of an angle $-t$, $t\in [0,s]$ about $u_0$ takes $(q,p)$ to the point $P^7_{(0,u_0)}(\ell_{s-t}, e^{i\tau}, r, \mu_{q^1}, \mu_{q^2},\lambda)$. Thus we obtain an arc of a circle parametrized by t, contained in $J^{-1}(0, u_0)\cap M_{G^4}$ and 
joining $(q,p)$ with the point $P^7_{(0,u_0)}(\ell_0, e^{i\tau}, r, \mu_{q^1}, \mu_{q^2},\lambda)$ which belongs to  $J^{-1}(0, u_0)\cap M_{G^4} \cap  \Pi_0$. This finishes the proof that
$J^{-1}(0, u_0)\cap M_{G^4}$ is arc-connected. Therefore the first equality of formula (\ref{optimo caso 7}) follows using Corollary \ref{corolario formula}. 

One can prove directly that $G_{\rho^7_{(0,u_0)}}$ has the stated expression. Besides, we know that $G_{\rho^7_{(0,u_0)}}$ acts properly and we can see directly using $P^7_{(0,u_0)}$ that the action is also free, therefore $\pi_{\rho^7_{(0,u_0)}}$ is a principal bundle. Now, we want to construct the orbit of the element $P^7_{(0,u_0)}(\ell_s,  e^{i\tau}, r, \mu_{q^1}, \mu_{q^2}, \lambda)$  under the action of $G_{\rho^7_{(0,u_0)}}$. A translation by $-\mu_{q^1}u_0$ takes this element to $P^7_{(0,u_0)}(\ell_s,e^{i\tau},r,0,\mu_{q^2}-\mu_{q^1},\lambda)$. By applying rotations about $u_0$ to this latter element one obtains a circle which intersects $\Pi_0$ in two points which identify the orbit, namely, $(\ell_0, e^{i[\tau]_{\pi}}, r, 0, \mu_{q^2}-\mu_{q^1}, \lambda)$ 
and $(\ell_0, e^{i([\tau]_{\pi}+\pi)}, r, 0, \mu_{q^2}-\mu_{q^1}, \lambda)$. Since the pair $\{e^{i[\tau]_{\pi}}, e^{i([\tau]_{\pi}+\pi)}\}$ is biunivocally determined by the element $\ell_{[\tau]_{\pi}}\in\mathcal{P}(\Pi_0)$ we have established a bijection
${\bf p}^7_{(0,u_0)}: \mathcal{P}(\Pi_0) \times\mathbb{R}^+ \times \mathbb{R}\times \mathbb{R} \rightarrow M_{\rho^7_{(0,u_0)}}$. Taking into account the previous construction of the space of orbits $M_{\rho^7_{(0,u_0)}}$ one can prove that $({\bf p}^7_{(0,u_0)})^{-1}\circ \pi_{\rho^7_{(0,u_0)}}$ is a submersion and we can deduce that ${\bf p}^7_{(0,u_0}$ is a diffeomorphism.

The rest of the proof can be performed in a a straightforward manner.  In particular, an explicit expression of the section is the following $(q,p)= \sigma_{\rho^7_{(0,u_0)}}\circ \bold{p}^7_{(0,u_0)}
(\ell_{\nu}^{\Pi_0}, r, \mu, \lambda)$ that is, 

\begin{equation}\label{17}\left\{\begin{array}{rcl}
q^1&=&\left((\lambda-1)\displaystyle\frac{r}{2}+(\lambda-1)\displaystyle\frac{r}{2}\cos(2\nu)-\mu\displaystyle\frac{r}{2}\sin(2\nu), 
\right.\vspace{.1cm}\\
&&\left.(\lambda -1)\displaystyle\frac{r}{2}sin(2\nu )-\mu\displaystyle\frac{r}{2}+\mu\displaystyle\frac{r}{2}\cos(2\nu), 0u_{03}\right)\vspace{.2cm}\\
q^2&=&\left(\lambda\displaystyle\frac{r}{2}+\lambda\frac{r}{2}\cos(2\nu)-\mu\displaystyle\frac{r}{2}\sin(2\nu),\right.\vspace{.1cm}\\
&&\left.\lambda\displaystyle\frac{r}{2}sin(2\nu )-\mu\displaystyle\frac{r}{2}+\mu\displaystyle\frac{r}{2}\cos(2\nu), \mu u_{03}\right)
\vspace{.2cm}\\
p^1&=&\left(\dfrac{r}{2}\sin(2\nu),\displaystyle\frac{r}{2}-\frac{r}{2}\cos(2\nu),\lambda u_{03}\right)\vspace{.2cm}\\
p^2&=&\left(-\displaystyle\frac{r}{2}\sin(2\nu),-\displaystyle\frac{r}{2}+\displaystyle\frac{r}{2}\cos(2\nu),(1-\lambda) u_{03}\right).
\end{array}\right.
\end{equation}

We will sometimes use an abuse of notation namely
$M_{\rho^7_{(0,u_0)}} \equiv \mathcal{P}(\Pi_0) \times \mathbb{R}^+\times \mathbb{R}
\times \mathbb{R}$ and then we will write
$\sigma_{\rho^7_{(0,u_0)}}(\ell^{\Pi_0}_{\nu}, r, \mu, \lambda)$ given by (\ref{17}). The reduced symplectic form $\omega_{\rho^7_{(0,u_0)}}$ is the pullback by $\sigma_{\rho^7_{(0,u_0)}}$ of the canonical symplectic form on $M$ and using (\ref{17}) one obtains the expression indicated in the statement. This is a long but straightforward calculation.

\ \\\emph{Proof of Case 8.}

In this case $J^{-1}(\delta_0 u_0,u_0)$ is the set of points
$(q^1,q^2,p^1,p^2)\in T^*(\mathbb{R}^3\times\mathbb{R}^3)$ that verify
\begin{eqnarray}
q^1\times p^1+q^2\times p^2&=&\delta_0 u_0.\label{preimagen de (al0,u0)paralelos1}\\
p^1+p^2&=&u_0.\label{preimagen de
(al0,u0)paralelos2}\end{eqnarray}

If $p^1\times p^2=0$ from (\ref{preimagen de
(al0,u0)paralelos2}) we obtain $p^1=\lambda u_0$ and $p^2=(1-\lambda)u_0$ with
$\lambda\in\mathbb{R}$. Then (\ref{preimagen de
(al0,u0)paralelos1}) becomes
$(\lambda q^1+(1-\lambda)q^2)\times u_0=\delta_0 u_0$, which is a contradiction. Therefore all solutions of (\ref{preimagen de
(al0,u0)paralelos1}) and  (\ref{preimagen de (al0,u0)paralelos2}) belong to $M_4$. In other words, the level set of the momentum map is contained in $M_4 \subseteq M_{G_4}$, then $J^{-1}(\delta_0 u_0,u_0)\cap M_{G_4}=J^{-1}(\delta_0 u_0,u_0)$. So, according to Corollary \ref{corolario formula}, if $J^{-1}(\delta_0 u_0,u_0)$ is connected then it must coincide with the level set of the optimal momentum map $\mathcal{J}^{-1}(\rho^8_{(\delta_0 u_0, u_0)})$. The connectedness of $J^{-1}(\delta_0 u_0,u_0)$ will be proven in a moment.

Since $N=p^1\times p^2\neq 0$, the  vectors $p^1, p^2, p^1-p^2$ and $u_0$ belong to $\Pi_N$.

Let $q^i=q_N^i+\tau^iN$, $i=1,2$, where $\tau^i\in\mathbb{R}$ and $q^i_N$
is perpendicular to $N$, then (\ref{preimagen de
(al0,u0)paralelos1}) and (\ref{preimagen de (al0,u0)paralelos2}) are equivalent to
\begin{eqnarray}
\tau^1N\times p^1+\tau^2N\times p^2&=&\delta_0 u_0,\label{preimagen de (al0,u0)paralelos-prima1}\\
q_N^1\times p^1+q_N^2\times p^2&=&0,\label{preimagen de (al0,u0)paralelos-prima2}
\\ p^1+p^2&=&u_0.\label{preimagen de (al0,u0)paralelos-prima3} \end{eqnarray}

Using (\ref{preimagen de (al0,u0)paralelos-prima1}) and (\ref{preimagen de (al0,u0)paralelos-prima3}) we can deduce that
\begin{equation}\left\{\begin{array}{rcl}\label{tau 1 y tau2 por separado}
\tau^1&=&\displaystyle\frac{\delta_0u_0^2}{N^2}-\displaystyle\frac{\delta_0}{N^2}<p^1,u_0>\vspace{.2cm}\\
\tau^2&=&-\displaystyle\frac{\delta_0}{N^2}<p^1,u_0>.
\end{array}\right.\end{equation}

and also that $\mid p^1\mid^2\mid u_0\mid^2-<p^1,u_0>^2= N^2$.
\ \\

Now we are going to parametrize $J^{-1}(\delta_0 u_0,u_0)$. Observe that equations (\ref{preimagen de (al0,u0)paralelos-prima2}), (\ref{preimagen de (al0,u0)paralelos-prima3}) are similar to the equations that appear in \emph{Case 7} namely $J(q_N, p) = (0, u_0)$ and then we can conclude that $(q_N, p) \in \mathcal{J}^{-1} (\rho^7_{(0, u_0)})$. Using the parametrization $P^7_{(0, u_0)}$ given by (\ref{optimo caso 7}) we can see that $N = p^1 \times p^2 = r_p \beta_s \times u_0$, therefore $r_p \neq 0$ and we must have $(q_N, p) \in W_{\mathfrak{p}}$ where $W_{\mathfrak{p}} \subseteq \mathcal{J}^{-1} (\rho^7_{(0, u_0)})$ is the chart described in \ref{Apendice C} which is connected. Since the equations $q^i = q^i_N + \tau^i N$, $i = 1,2$ give a $C^{\infty}$ map $\mathcal{F}_{\mathfrak{p}}: W_{\mathfrak{p}}\rightarrow M$ whose image is $J^{-1}(\delta_0u_0, u_0)$, we have that $J^{-1}(\delta_0u_0, u_0)$ is connected. We can conclude, using an abuse of notation , that $\mathcal{F}_{\mathfrak{p}} : W_{\mathfrak{p}} \rightarrow \mathcal{J}^{-1} (\rho^8_{(\delta_0u_0, u_0)})$ is a $C^{\infty}$ surjective map.
One can deduce from the definition that $\mathcal{F}_{\mathfrak{p}}$ is injective and  that
$\mathcal{F}_{\mathfrak{p}}^{-1}$ is a $C^{\infty}$ map which shows that 
$\mathcal{F}_{\mathfrak{p}}$ is a diffeomorphism.

If $A \in SO(3)$ satisfies $Au_0 = u_0$ then $A$ leaves $\tau^1$ and $\tau^2$ invariant. A group element $(A, a)$ that leaves equation (\ref{preimagen de (al0,u0)paralelos-prima3}) invariant must satisfy $Au_0 = u_0$ then $A$ leaves $\tau^1$ and $\tau^2$ invariant. If, in addition, $(A, a)$ leaves equation (\ref{preimagen de (al0,u0)paralelos-prima2}) invariant then  $a = \xi u_0$ for some $\xi \in \mathbb{R}$.
Since $\mathcal{J}^{-1} (\rho^8_{(\delta_0 u_0, u_0)})$ is defined by
(\ref{preimagen de (al0,u0)paralelos-prima1}), (\ref{preimagen de (al0,u0)paralelos-prima2}) and (\ref{preimagen de (al0,u0)paralelos-prima3}) we can conclude that $G_{\rho^7_{(0, u_0)}}=G_{\rho^8_{(\delta_0 u_0 , u_0)}}$. Moreover one can check that $\mathcal{F}_{\mathfrak{p}}$ is equivariant under the action of this group and we can deduce that there exists a diffeomorphism $\mathcal{F}_{\mathfrak{p}, \rho^7_{(0,u_0)}}$ satisfying (\ref{formula Fp}).

Therefore the set of points $(q_N,p)$ such that $(q,p)\in J^{-1}(\delta_0 u_0,u_0)$ where $q^1=q^1_N+\tau^1 N$, $q^2=q^2_N+\tau^2N$ must have a parametrization similar to the one that we have in  \emph{Case 7} namely

\begin{equation}\label{parametrica en cada s-2}\left\{\begin{array}{rcl}
q^1_N&=&r_{q^1_N}\beta_s+\mu_{q^1_N}u_0,\ \ \ r_{q^1_N},\mu_{q^1_N}\in\mathbb{R}\vspace{.2cm}\\
q^2_N&=&(r_{q^1_N}+r_{q^1_Nq^2_N})\beta_s+\mu_{q^2_N}u_0,\ \ \
r_{q^1_Nq^2_N},\mu_{q^2_N}\in\mathbb{R}\vspace{.2cm}\\
p^1&=&r_p\beta_s+\lambda u_0,\ \ \ \lambda,r_p\in\mathbb{R}\vspace{.2cm}\\
p^2&=&-r_p\beta_s+(1-\lambda)u_0,
\end{array}\right.\end{equation} with the conditions
\begin{equation}\label{condicion 1-2}
r_p\neq 0,
\end{equation}
\begin{equation}\label{condicion 2-2}
r_{q^1_N}+r_{q^1_Nq^2_N}(1-\lambda)+r_p(\mu_{q^2_N}-\mu_{q^1_N})=0.
\end{equation}

Using (\ref{tau 1 y tau2 por separado}), (\ref{parametrica en cada s-2}), (\ref{condicion 1-2}) and (\ref{condicion 2-2}) we can deduce after some work
\begin{equation}\left\{\begin{array}{rcl}\label{tau 1 y tau2 por separado 2}
\tau^1&=&\displaystyle\frac{\delta_0u_0^2}{r_p^2u_0^2}-\displaystyle\frac{\delta_0}{r_p^2u_0^2}<p^1,u_0>\vspace{.2cm}\\
\tau^2&=&-\displaystyle\frac{\delta_0}{r_p^2u_0^2}<p^1,u_0>.
\end{array}\right.
\end{equation}
Finally, the formula (\ref{tau 1 y tau2 por separado 2}) can be transformed as follows
\begin{equation}\left\{\begin{array}{rcl}\label{tau 1 y tau2 por separado 4}
\tau^1&=&\displaystyle\frac{\delta_0}{r_p^2}(1-\lambda)\vspace{.2cm}\\
\tau^2&=&-\displaystyle\frac{\delta_0\lambda}{r_p^2}.
\end{array}\right.
\end{equation}

In order to simplify the proof of this transformation, we can take w.l.g that $u_0=(0,0,u_{03})$ and therefore $u_0^2=u_{03}^2$, $<p^1,u_0>=\lambda u_{03}^2$. 

Collecting results we obtain the following parametrization of $J^{-1}(\delta_0 u_0,u_0)$ 
\begin{equation}\label{parametrica en cada s-3}\left\{\begin{array}{rcl}
q^1&=&r_{q^1_N}\beta_s+\mu_{q^1_N}u_0+\displaystyle\frac{\delta_0}{r_p}(1-\lambda)\beta_s\times u_0,\ \ \ r_{q^1_N},\mu_{q^1_N}\in\mathbb{R}\vspace{.2cm}\\
q^2&=&(r_{q^1_N}+r_{q^1_Nq^2_N})\beta_s+\mu_{q^2_N}u_0-\displaystyle\frac{\delta_0\lambda}{r_p}\beta_s\times u_0,\ \ \
r_{q^1_Nq^2_N},\mu_{q^2_N}\in\mathbb{R}\vspace{.2cm}\\
p^1&=&r_p\beta_s+\lambda u_0,\ \ \ \lambda,r_p\in\mathbb{R}\vspace{.2cm}\\
p^2&=&-r_p\beta_s+(1-\lambda)u_0,
\end{array}\right.
\end{equation} where the conditions
$r_p\neq 0$ and $r_{q^1_N}+r_{q^1_Nq^2_N}(1-\lambda)+r_p(\mu_{q^2_N}-\mu_{q^1_N})=0$ and $s\in [0,2\pi)$ must be satisfied.
So $J^{-1}(\delta_0u_0, u_0)=\mathcal{J}^{-1}(\rho^8_{(\delta_0 u_0, u_0)}$ is a $6$-dimensional manifold diffeomorphic to the chart $W_{\mathfrak{p}}$ of \emph{Case 7}.

Next we will prove (\ref{omega 8}). The first equality follows from the definition of $\sigma_{\rho^8_{(\delta_0 u_0, u_0)}}$ and the last equality was proven in \emph{Case 7}. 

The second equality follows using the last one and a long but straightforward calculation of the pullback appearing in the r.h.s of the first equality, taking into account the definition of $\mathcal{F}_{\mathfrak{p}}$. It is convenient to proceed as follows. If in equation (\ref{17}) the symbols $q^1$ and  $q^2$ are replaced by  $q^1_N$ and $q^2_N$ respectively one obtains the expression of the value of $(q^1_N, q^2_N, p^1, p^2) = \sigma_{\rho^7_{(0, u_0)}}\circ {\bf p}^7_{(0, u_0)}$  at the point $(\ell^{\Pi_0}_{\nu_N}, r, \mu_N, \lambda)$.  Similarly, one obtains the expression of $\tau^1N$ and 
$\tau^2N$ as a function of $(\ell^{\Pi_0}_{\nu_N}, r, \mu_N, \lambda)$, which will be called $\Delta  q^1$, $\Delta  q^2$, namely
\begin{equation}\label{deltas}
\begin{array}{rcl}
\Delta q^1 &=& \left((\delta_0 u_{03}\frac{(1-\lambda)}{r},-\delta_0 u_{03}\frac{(1-\lambda)}{r}cotg(\nu_N),0 \right)\\
\Delta q^2 &=& \left(-\delta_0 u_{03}\frac{\lambda}{r},\delta_0 u_{03}\frac{\lambda}{r}cotg(\nu_N),0 \right). 
\end{array}
\end{equation}
We will sometimes use an abuse of notation namely
$M_{\rho^8_{(\delta_0 u_0, u_0)}} \equiv \left(\mathcal{P}(\Pi_0)-
\{\ell^{\Pi_0}_0\} \right)
\times \mathbb{R}^+\times \mathbb{R}\times \mathbb{R} \equiv
(0,\pi) \times \mathbb{R}^+\times \mathbb{R}\times \mathbb{R}$ and then we can write

\begin{equation}\label{sigma8}
\sigma_{\rho^8_{(\delta_0 u_0, u_0)}}(\ell^{\Pi_0}_{\nu_N}, r, \mu_N, \lambda)
=\left(\sigma_{\rho^7_{(0, u_0)}}+(\Delta q^1,\Delta q^2,0,0)\right)(\ell^{\Pi_0}_{\nu_N}, r, \mu_N, \lambda).
\end{equation}

Using this we can deduce that the value of the r.h.s of the first equality of (\ref{omega 8}) at the point $(\ell^{\Pi_0}_{\nu_N}, r, \mu_N, 
\lambda)$ can be written
\begin{equation*}\begin{array}{rcl}
&\displaystyle\sum_{i=1}^{3}(dq^1_i + d \Delta  q^1_i)
(\ell^{\Pi_0}_{\nu_N}, r, \mu_N, \lambda)
\wedge d p^1_i (\ell^{\Pi_0}_{\nu_N}, r, \mu_N, 
\lambda)\\&+\displaystyle\sum_{i=1}^{3}(dq^2_i + d \Delta  q^2_i)
(\ell^{\Pi_0}_{\nu_N}, r, \mu_N, \lambda)\wedge 
d p^2_i (\ell^{\Pi_0}_{\nu_N}, r, \mu_N, \lambda).
\end{array}
\end{equation*}

A straightforward calculation shows that
$$\displaystyle\sum_{i=1}^{3}d \Delta  q^1_i (\ell^{\Pi_0}_{\nu_N}, r, \mu_N, \lambda)\wedge d p^1_i (\ell^{\Pi_0}_{\nu_N}, r, \mu_N, \lambda)+\displaystyle\sum_{i=1}^{3}
d \Delta  q^2_i (\ell^{\Pi_0}_{\nu_N}, r, \mu_N, \lambda)
\wedge d p^2_i (\ell^{\Pi_0}_{\nu_N}, r, \mu_N, 
\lambda) = 0,$$
and so we have proven the second equality of (\ref{omega 8}).

Now, using (\ref{omega 8}) and the fact that $\omega_{\rho^8_{(\delta_0 u_0, u_0)}}=\sigma_{\rho^8_{(\delta_0 u_0, u_0)}}^*\Omega_M$ and $\omega_{\rho^7_{(0, u_0)}}=\sigma_{\rho^7_{(0, u_0)}}^*\Omega_M$ we can deduce that $(\mathcal{F}_{\mathfrak{p},\rho^7_{(0, u_0)}})^*\omega_{\rho^8_{(\delta_0 u_0, u_0)}}=\omega_{\rho^7_{(0, u_0)}}\vert_{\pi_{\rho^7_{(0,u_0)}}(W_{\mathfrak{p})}}$. So $\mathcal{F}_{\mathfrak{p},\rho^7_{(0, u_0)}}$ is a simplectomorfhism. Finally using  (\ref{formula Fp}) and the fact that the presymplectic forms $\omega_{W_{\mathfrak{p}}}$ and $\omega_{\mathcal{J}^{-1}(\rho^8_{(\delta_0 u_0, u_0)})}$ coincide with $(\pi_{\rho^7_{(0, u_0)})}\vert_{W_{\mathfrak{p}}})^*(\omega_{\rho^7_{(0, u_0)}}\vert_{\pi_{\rho^7_{(0, u_0)}}(W_{\mathfrak{p}})})$ and $(\pi_{\rho^8_{(\delta_0 u_0, u_0)}})^*(\omega_{\rho^8_{(\delta_0 u_0, u_0)}})$ respectively, we can deduce tha $\mathcal{F}_{\mathfrak{p}}$ is a presymplectomorphism. The proof of the stated trivialization of $\pi_{\rho^8_{(\delta_0 u_0, u_0)}}$ offers no difficulty and will be omitted.



\ \\\emph{Proof of Case 9.}

Since $(I,a^9_{(\alpha_0, u_0)})J^{-1}(0,u_0)=J^{-1}(\alpha_0,u_0)$ and $(I,a^9_{(\alpha_0, u_0)})M_{G^4}= M_{G^4}$ we have $(I,a^9_{(\alpha_0, u_0)})\mathcal{J}^{-1}(\rho^7_{(0,u_0)})=(I,a^9_{(\alpha_0, u_0)})(J^{-1}(0,u_0)\cap M_{G^4})=J^{-1}(\alpha_0,u_0)\cap M_{G^4}$. Then $J^{-1}(\alpha_0,u_0)\cap M_{G^4}$ is connected and we can deduce that $\mathcal{J}^{-1}(\rho^9_{(\alpha_0, u_0)})=(I,a^9_{(\alpha_0, u_0)})\mathcal{J}^{-1}(\rho^7_{(0,u_0)})=J^{-1}(\alpha_0,u_0)\cap M_{G^4}$. The rest of the proof follows without difficulty.

By an abuse of notation we can identify
$M_{\rho^9_{(\alpha_0, u_0)}} \equiv \mathcal{P}(\Pi_0) \times \mathbb{R}^+\times \mathbb{R}
\times \mathbb{R}$ and then we have that
\begin{equation}\label{sigma9}
\sigma_{\rho^9_{(\alpha_0, u_0)}}(\ell^{\Pi_0}_{\nu}, r, \mu, \lambda)
= \sigma_{\rho^7_{(0,u_0)}}(\ell^{\Pi_0}_{\nu}, r, \mu, \lambda) +
(a^9_{(\alpha_0, u_0)}, a^9_{(\alpha_0, u_0)},0,0).
\end{equation}

\ \\\emph{Proof of Case 10.}
 
Since $(I,a^{10}_{(\alpha_0, u_0)})J^{-1}(\delta^{10}_{(\alpha_0, u_0)}u_0, u_0)=J^{-1}(\alpha_0, u_0)$ and $(I,a^{10}_{(\alpha_0, u_0)})M_{G^4}= M_{G^4}$ we have $(I,a^{10}_{(\alpha_0, u_0)})\mathcal{J}^{-1}(\rho^8_{(\delta^{10}_{(\alpha_0, u_0)}u_0,u_0)})=(I,a^{10}_{(\alpha_0, u_0)})(J^{-1}(\delta^{10}_{(\alpha_0, u_0)}u_0,u_0)\cap M_{G^4})=J^{-1}(\alpha_0,u_0)\cap M_{G^4}$. Then $J^{-1} (\alpha_0,u_0)\cap M_{G^4}$ is connected and therefore
$\mathcal{J}^{-1}(\rho^{10}_{(\alpha_0,u_0)})=(I,a^{10}_{(\alpha_0, u_0)})\mathcal{J}^{-1}(\rho^8_{(\delta^{10}_{(\alpha_0, u_0)}u_0,u_0)})=J^{-1}(\alpha_0,u_0)\cap M_{G^4}$. The rest of the proof follows without difficulty.

By an abuse of notation we can identify
$M_{\rho^{10}_{(\alpha_0,u_0)}} \equiv 
\left(\mathcal{P}(\Pi_0)- \{\ell^{\Pi_0}_0 \}\right)\times \mathbb{R}^+\times \mathbb{R}
\times \mathbb{R}$ and then we have that
\begin{equation}\label{sigma10}
\sigma_{\rho^{10}_{(\alpha_0,u_0)}}(\ell^{\Pi_0}_{\nu}, r, \mu, \lambda)=\sigma_{\rho^8_{(\delta^{10}_{(\alpha_0, u_0)}u_0,u_0)}}(\ell^{\Pi_0}_{\nu}, r, \mu, \lambda) +
(a^{10}_{(\alpha_0, u_0)}, a^{10}_{(\alpha_0, u_0)},0,0).
\end{equation}

\boldparagraph{\emph{Proof of (B)}}

(B) follows directly from (A).

\section{Summary and conclusions}\label{sec:prel}

Some useful information from the $10$ cases of Theorem \ref{Teorema principal}  will be conveniently collected in a Table. The image of $\Xi$, which is the set of all $((\alpha,u),S)$ such that $J^{-1}(\alpha,u)\cap M_S\neq \emptyset$, is biunivocally related to  the $10$ columns of the Table where, for each column, $(\alpha,u)$ and $S$ appear in the first and second row, respectively and the rest of the rows contain some information about the optimal reduction process, like $\mathcal{J}^{-1}(\rho)$, $G_{\rho}$, $M_{\rho}$ and $\omega_{\rho}$. The map $\pi_{\rho}: \mathcal{J}^{-1}(\rho) \rightarrow M_{\rho}$ is a trivial principal bundle with group $G_{\rho}$ in \emph{Cases 6, 7, 8, 9, 10}, therefore, $\mathcal{J}^{-1}(\rho)$ is isomorphic to $G_{\rho} \times M_{\rho}$ in those cases. Sections giving those trivializations are, respectively,
$\sigma^{\bf f}_{\rho^6_{(\alpha_0,0)}}$, (\ref{sigma6}); $\sigma_{\rho^7_{(0,u_0)}}$, (\ref{17}); $\sigma_{\rho^8_{(\delta_0 u_0, u_0)}}$, 
(\ref{sigma8}); $\sigma_{\rho^9_{(\alpha_0, u_0)}}$, (\ref{sigma9}); $\sigma_{\rho^{10}_{(\alpha_0,u_0)}}$, (\ref{sigma10}). For a given $SE(3)$ invariant Hamiltonian on $M$ the reduced Hamiltonian can be obtained by a pullback using those sections.
In \emph{Cases 1, 2, 3, 4, 5} the action of $G_{\rho}$ on $\mathcal{J}^{-1}(\rho)$ is not free. The maps $\pi_{\rho^3_{(0,\delta_0 y_0; y_0)}}$ and $\pi_{\rho^5_{(\delta_0 z_0, \delta_0 y_0;x_0, y_0)}}$ are trivial principal bundles with group $\mathcal{T}_{y_0}$ and with trivializing sections $\sigma_{\rho^3_{(0,\delta_0 y_0; y_0)}}$, (\ref{sigma3})  and  
$\sigma_{\rho^5_{(\delta_0 z_0, \delta_0 y_0;x_0, y_0)}}$, (\ref{sigma5}), respectively. For a given $SE(3)$ invariant Hamiltonian on $M$ the reduced Hamiltonian can be obtained by a pullback using those sections. The optimal reduced spaces $M_{\rho^3_{(0,\delta_0 y_0; y_0)}}$ and $M_{\rho^5_{(\delta_0 z_0, \delta_0 y_0;x_0, y_0)}}$ are diffeomorphic to $\mathbb{R}^2$.

By definition $M^+_{\rho^2_{(0,0;y_0)}} := \mathcal{J}^{-1}(\rho^2_{(0,0;y_0)})/_{G^+_{\rho^2_{(0,0;y_0)}}}$ with $G^+_{\rho^2_{(0,0;y_0)}}:= G^2(y_0) \circledS \mathcal{T}_{y_0}$. Similarly, $M^+_{\rho^4_{(0,0;x_0, y_0)}} := \mathcal{J}^{-1}(\rho^4_{(0,0;x_0, y_0)})/_{G^+_{\rho^4_{(0,0;x_0, y_0)}}}$ with
$G^+_{\rho^4_{(0,0;x_0, y_0)}}:= G^3(x_0, y_0) \circledS \mathcal{T}_{y_0} $

The maps $\pi_{\rho^2_{(0,0;y_0)}}$ and $\pi_{\rho^4_{(0,0;x_0, y_0)}}$ are nontrivial principal bundles. They are compositions $\pi_{\rho^2_{(0,0;y_0)}} = D^2\circ \pi^+_{\rho^2_{(0,0;y_0)}}$ and $\pi_{\rho^4_{(0,0;x_0, y_0)}} = D^4\circ  \pi^+_{\rho^4_{(0,0;x_0, y_0)}}$.  
 
The maps
$\pi^+_{\rho^2_{(0,0;y_0)}}: \mathcal{J}^{-1} (\rho^2_{(0,0;y_0)})\rightarrow M^+_{\rho^2_{(0,0;y_0)}}$ and  
\newline $\pi^+_{\rho^4_{(0,0;x_0,y_0)}}:\mathcal{J}^{-1}(\rho^4_ {(0,0;x_0,y_0)})\rightarrow M^+_{\rho^4_{(0,0;x_0, y_0)}}$  are trivial principal bundles isomorphic to 
$\mathcal{T}_{y_0} \times M^+_{\rho^2_{(0,0;y_0)}}$ and
$\mathcal{T}_{y_0} \times M^+_{\rho^4_{(0,0;x_0, y_0)}}$ with trivializing sections $\sigma^+_{\rho^2_{(0,0;y_0)}}$, (\ref{sigma2}) and $\sigma^+_{\rho^4_{(0,0;x_0, y_0)}}$, (\ref{sigma4}), respectively. For a given $SE(3)$ invariant Hamiltonian on $M$ the reduced Hamiltonians on $M^+_{\rho^2_{(0,0;y_0)}}$ or
$M^+_{\rho^4_{(0,0;x_0, y_0)}}$ for the first stage can be obtained by a pullback using those sections.
The maps  $D^2:M^+_{\rho^2_{(0,0;y_0)}} \rightarrow M_{\rho^2_ {(0,0;y_0)}}$ and $D^4:M^+_{\rho^4_ {(0,0;x_0, y_0)}} \longrightarrow M_{\rho^4_{(0,0;x_0, y_0)}}$ are double coverings and the reduced Hamiltonians on $M_{\rho^2_{(0,0;y_0)}}$ or $M_{\rho^4_{(0,0;x_0, y_0)}}$ are obtained by passing to the quotient by $D^2$ or $D^4$. The manifolds $M^+_{\rho^2_{(0,0;y_0)}}$, $M_{\rho^2_{(0,0;y_0)}}$, $M^+_{\rho^4_{(0,0;x_0, y_0)}}$ and $M_{\rho^4_{(0,0;x_0, y_0)}}$ are diffeomorphic to $\mathbb{R}^2-\{0\}$. 

The Table is divided in two  parts. The first one contains the \emph{Cases 1} to \emph{5} and the second one contains the \emph{Cases 6} to \emph{10}. In order to shorten the notation in the Table we have used sometimes the notation $\rho^k$ instead of $\rho^k_{subscript}$ and also $a^k$
instead of $a^k_{subscript}$.

\begin{center}\small
\begin{tabular}{|c|c|c|c|c|c|}
\hline
&1&2&3&4&5\\
\hline
&&&&&\\
$(\alpha_0, u_0),$&$(0,0)$&$(0,0)$&$(0,\delta_0y_0), \delta_0\neq0$&$(0,0)$&$(\delta_0z_0,\delta_0 y_0)$,$\delta_0\neq 0$\\
\small value of $J$&&&$y_0\neq0$&&$ z_0=x_0\times y_0\neq 0$\\
\hline
&&&&&\\
$S$, isotropy&$G^1(x_0)$&$G^2(y_0), y_0\neq 0$&$G^2(y_0), y_0\neq 0$&$G^3(x_0,y_0)$&$G^3(x_0,y_0)$\\
group of $T^*\phi$&&&&$x_0\times y_0\neq 0$&$x_0\times y_0\neq 0$\\
\hline
&&&&&\\
$\rho$, value&$\rho^1_{(0,0;x_0)}$&$\rho^2_{(0,0; y_0)}$&$\rho^3_{(0,\delta_0 y_0; y_0)}$&$\rho^4_{(0,0;x_0,y_0)}$&$\rho^5_{(\delta_0 z_0, \delta_0 y_0;x_0, y_0)}$\\
of $\mathcal{J}$&&&&&$z_0=x_0 \times y_0$\\
\hline
&&&&&\\
$\mathcal{J}^{-1}(\rho)$&$\{(x_0,x_0,0,0)\}$&$\mathcal{T}_{y_0}\times M^+_{\rho^2}$&$\mathcal{T}_{y_0}\times M_{\rho^3}$&$\mathcal{T}_{y_0}\times M^+_{\rho^4}$&$\mathcal{T}_{y_0}\times M_{\rho^5}$\\
Dimension&0&3&3&3&3\\
\hline
$G_{\rho}$, optimal&&&&&\\
isotropy group&$G^1(x_0)$&$\mathbb{Z}^{(2)}_2 \circledS G^+_{\rho^2}$&$G^2(y_0) \circledS \mathcal{T}_{y_0}$&$\mathbb{Z}^{(4)}_2\circledS G^+_{\rho^4}$&$G^3(x_0, y_0) \circledS \mathcal{T}_{y_0}$\\
&&&&&\\
\hline
$M_{\rho}$, Optimal&&&&&\\
reduced space&$\{[(x_0,x_0,0,0)]\}$&$\mathbb{R}^2-\{0\}$&$\mathbb{R}^2$&$\mathbb{R}^2-\{0\}$&$\mathbb{R}^2$\\
Dimension&0&2&2&2&2\\
\hline
Optimal reduced&&&&&\\
symplectic form&Trivial&$\displaystyle\frac{1}{2}|y_0|^2rdr\wedge d\varphi$&$|y_0|^2d\delta\wedge d\gamma$&$\displaystyle\frac{1}{2}|y_0|^2rdr\wedge d\varphi$&$|y_0|^2d\delta\wedge d\gamma$\\
&&&&&\\
\hline
\end{tabular}
\end{center}

\ \\

\begin{center}\small
\begin{tabular}{|c|c|c|c|c|c|}
\hline
&6&7&8&9&10\\
\hline
&&&&$(\alpha_0,u_0)$&$(\alpha_0,u_0)$\\
$(\alpha_0, u_0),$&$(\alpha_0, 0)$&$(0,u_0)$&$(\delta_0u_0,u_0)$&$\alpha_0\times u_0\neq 0$&$\alpha_0\times u_0\neq 0$\\
value of $J$&$\alpha_0\neq 0$&$u_0\neq 0$&$\delta_0\neq0, u_0\neq 0$&$<\alpha_0,u_0>=0$&$<\alpha_0,u_0>\neq 0$\\
\hline
&&&&&\\
$S$, isotropy&$G^4$&$G^4$&$G^4$&$G^4$&$G^4$\\
group of $T^*\phi$&&&&&\\
\hline
&&&&&\\
$\rho$, value&$\rho^6_{(\alpha_0, 0)}$&$\rho^7_{(0,u_0)}$&$\rho^8_{(\delta_0 u_0, u_0)}$&$\rho^9_{(\alpha_0, u_0)}$&$\rho^{10}_{(\alpha_0, u_0)}$\\
of $\mathcal{J}$&&&&&\\
\hline
&&&&&\\
$\mathcal{J}^{-1}(\rho)$&$G_{\rho}\times M_{\rho}$&$G_{\rho}\times M_{\rho}$&$G_{\rho}\times M_{\rho}$&$G_{\rho}\times M_{\rho}$&$G_{\rho}\times M_{\rho}$\\
Dimension&6&6&6&6&6\\
\hline
$G_{\rho}$, optimal&&&&&\\
isotropy group&$G^2(\alpha_0)\circledS\mathbb{R}^3$&$G^2(u_0) \circledS \mathcal{T}_{u_0}$&$G^2(u_0) \circledS \mathcal{T}_{u_0}$&${\bf C}_{a^9}(G_{\rho^7})$&${\bf C}_{a^{10}_{(\alpha_0, u_0)}}(G_{\rho^8})$\\
&&&&$a^9\equiv a^9_{(\alpha_0, u_0)}$&$\rho^8\equiv\rho^8_{(\delta^{10}_{(\alpha_0, u_0)} u_0, u_0)})$\\
\hline
$M_{\rho}$, Optimal&&&&&\\
reduced space&$\mathbb{R}^+\times\mathbb{R}$&$\mathbb{R}P^1\times\mathbb{R}^+\times\mathbb{R}^2$&$(0,\pi)\times\mathbb{R}^+\times\mathbb{R}^2$&$\mathbb{R}P^1\times\mathbb{R}^+\times\mathbb{R}^2$&$(0,\pi)\times\mathbb{R}^+\times\mathbb{R}^2$\\
Dimension&2&4&4&4&4\\
\hline
&&&&&\\
Optimal reduced&$cd\lambda\wedge dc$&$-rdr \wedge d\nu$&$-r_Ndr_N \wedge d\nu_N$&$-rdr \wedge d\nu$&$-r_Ndr_N \wedge d\nu_N$\\
symplectic form&&$-u ^2_{03} d\mu \wedge d\lambda$&$-u ^2_{03} d\mu_N \wedge d\lambda$&$-u ^2_{03} d\mu \wedge d\lambda$&$-u ^2_{03} d\mu_N \wedge d\lambda$\\
&&&&&\\
\hline
\end{tabular}
\end{center}

\boldparagraph{Comparison between Marsden-Weinstein reduction and optimal reduction.} See \cite{O-R} for a comparison between symplectic and optimal reduction in general. We will use the notation introduced in Subsection \ref{3.1}. Using formula (\ref{j como union de optimos}) with $\mu = (\alpha, u)$, we see that $Im(\Xi)$ is the set of all  $((\alpha,u),S)$ such that  $(\alpha,u)$ and $S$ are the first and second element of some column of Table. In order to determine for each $(\alpha,u)$ the set of all the $\mathcal{J}^{-1}(\rho_{((\alpha,u),S)})$ appearing in the disjoint union (\ref{j como union de optimos}), we should simply find in Table all the columns where $(\alpha,u)$ appears as the first element and then determine the set of all the $S$ that appear as the second element in one of each such columns. 

For instance, choose $(0,u_0)$ ), $u_0 \neq 0$, as being the value of the momentum map. Such element appears only as the first element of column $3$ with $y_0=u_0/\delta_0$, and of column $7$, therefore the set of all $S$ that appear as the second element of column $3$  and column $7$ is $\{G^2(y_0), G^4\}$. So we obtain 
\begin{equation}\label{J-1(0,u0)}
\begin{array}{rcl}
J^{-1}(0,u_0)&=&\mathcal{J}^{-1}(\rho_{((0,u_0),G^2(y_0))})\cup 
\mathcal{J}^{-1}(\rho_{((0,u_0),G^4)}\vspace{.2cm}\\
&=&\mathcal{J}^{-1}(\rho^3_{(0,u_0;[y_0])})\cup\mathcal{J}^{-1}(\rho^7_{(0,u_0)}).
\end{array}
\end{equation} 

The momentum mapping  $(\alpha, u) = J(q,p)$ is defined by equations (\ref{preimagen del momento alpha}), (\ref{preimagen del momento u}), and one can prove by elementary calculations that the tangent map  $T_{(q,p)}J$ is surjective if and only if the condition 
\begin{equation}\label{N1}
q^1 \parallel q^2 \parallel p^1 \parallel p^2  
\end{equation}
is not satisfied.  

According to formula (\ref{optimo caso 7}) in Case 7, elements $(q,p) \in \mathcal{J}^{-1}(\rho^7_{(0, u_0)})$ satisfy
$r_{q^1q^2}^2 + r_{p^2}^2 >0$, from which one can deduce that
condition (\ref{N1}) is not satisfied and therefore
$T_{(q,p)}J$ is surjective at all points of
$\mathcal{J}^{-1}(\rho^7_{(0, u_0)})$. On the other hand, from formula (\ref{Caso 3}) in Case 3, we see that if $(q,p) \in \mathcal{J}^{-1}(\rho^3_{(0, u_0;y_0)})$ then condition (\ref{N1}) is satisfied and therefore $T_{(q,p)}J$ is not surjective at those points. We can conclude that $\mathcal{J}^{-1}(\rho^7_{(0, u_0)})$ coincides with the set of regular points of $J$. Besides, the momentum isotropy subgroup $G_{(0,u_0)}$ and the optimal momentum isotropy subgroup $G_{\rho^7_{(0, u_0)}}$ coincide and then the optimal  reduction bundle and the symplectic reduction bundle on the set of regular points coincide.

Now, we would like  to mention two things. First, symplectic reduction does not give important additional information about the reduction bundle. But, our results in Case 7 inspired by optimal reduction, tell us that this reduction bundle is a trivial principal bundle and, moreover, we have obtained an explicit section (see formula (\ref{17}), whose geometric insight appears in the proof of Case 7). This is useful in concrete applications for instance in order to find the reduced Hamiltonian by a pullback using the section. Second, symplectic reduction does not say anything about what to do with reduction on the set of points in the complement in $J^{-1}(0,u_0)$ of the set of regular points of $J$, in fact,  the action of $G_{(0,u_0)}$ on the complemnt is not free so symplectic reduction cannot be applied. However,  optimal reduction tells us that this complement is precisely the level set of the optimal momentum of Case 3, so one knows how to do optimal reduction also on the complement.

Now, let us consider the case obtained by conjugation from
(\ref{J-1(0,u0)}). From \emph{Case 9} we know that $T_{a^9_{(\alpha_0, u_0)}} J^{-1} (0, u_0)=J^{-1} (\alpha_0, u_0)$.  On the other hand, if we set in formula (\ref{relacion caso 5 y 3})  of \emph{Case 5}, $x_0=a^9_{(\alpha_0, u_0)}$, $y_0 = u_0$, $\delta_0=1$ and $z_0 = \alpha_0$ we obtain 
$T_{a^9_{(\alpha_0, u_0)}}\mathcal{J}^{-1}(\rho^3_{(0,u_0;u_0)})=\mathcal{J}^{-1}(\rho^5_{(\alpha_0, u_0;a^9_{(\alpha_0 , u_0)}, u_0)})$. We can conclude that, for $(\alpha_0, u_0)$ such that $\alpha_0 \times u_0 \neq 0$ and $<\alpha_0, u_0> = 0$ we have 
\begin{equation}\label{F4}J^{-1} (\alpha_0, u_0) = 
\mathcal{J}^{-1}(\rho^9_{(\alpha_0, u_0)} \cup 
\mathcal{J}^{-1}(\rho^5_{(\alpha_0, u_0; 
a^9_{(\alpha_0 , u_0)}, u_0)});\end{equation}

One can prove, using the parametrizations of Case 9 and Case 5 that the condition (\ref{N1}) is not satisfied for all points of $J^{-1}(\alpha_0, u_0)$ so they are regular points of $J$. Also using conjugation one can deduce that the optimal momentum isotropy subgroup $G_{\rho^9_{(\alpha_0, u_0)}}$ and the momentum isotropy subgroup $G_{(\alpha_0, u_0)}$ coincide and, besides, that $G_{(\alpha_0, u_0)}$ acts freely on 
$\mathcal{J}^{-1}(\rho^9_{(\alpha_0, u_0)})$ but not freely
on $\mathcal{J}^{-1}\left(\rho^5_{(\alpha_0, u_0;a^9, u_0)}\right)$, so we see that symplectic reduction cannot be applied. 
\ \\

By applying an entirely similar procedure we obtain the following disjoint union 
\begin{equation}\label{F1}J^{-1}(0,0)=\displaystyle\bigcup_{x_0\in\mathbb{R}^3}\mathcal{J}^{-1}(\rho^1 (x_0))\cup\displaystyle\bigcup_{[y_0]\in P^2}\mathcal{J}^{-1}(\rho^2_{(0,0;[y_0])})\cup\displaystyle\bigcup_{[x_0, y_0]\in P^3}\mathcal{J}^{-1} (\rho^4_{(0,0;[x_0 , y_0])}).\end{equation}

For short we will sometimes denote the unions over
$x_0$, $[y_0]$, $[x_0, y_0]$ in (\ref{F1}) as, respectively,
$U_1$, $U_2$, $U_3$.

The set of points of $J^{-1}(0,0)$ where $J$ is regular coincides with $U_3$.  On the other hand the momentum isotropy subgroup $G_{(0,0)}$ coincides with $SE(3)$,  which acts transitively on  $U_2 \cup  U_3$. This action is not free and besides, $G_{(0,0)}$ does not act on $U_3$ or $U_2$ separately. On the other hand, $G_{(0,0)}$ acts transitively on $U_1$ but not freely. As far as we know, freeness of the action is a requirement, by definition, for symplectic reduction, so no comparison is possible with optimal reduction in this case.

By the same method we can deduce that

\begin{equation}\label{F2}J^{-1}(\alpha_0,0)=\mathcal{J}^{-1}(\rho^6_{(\alpha_0,0)});\end{equation} 
\begin{equation}\label{F3}J^{-1}(\delta_0 u_0, u_0)=\mathcal{J}^{-1}(\rho^8_{(\delta_0 u_0, u_0)}).\end{equation}

Finally, we can deduce from \emph{Case 10} that 
\begin{equation}\label{F5}\mathcal{J}^{-1} (\rho^{10}_{(\alpha_0,u_0)})=J^{-1} (\alpha_0, u_0).\end{equation}

From (\ref{F2}), (\ref{F3}), (\ref{F5}),  we can deduce that symplectic reduction and optimal reduction coincide in \emph{Cases 6, 8} and \emph{10}. 

We have considered all the possibilities of comparison between synplectic reduction and optimal reduction for the $2$-Body problem.
\ \\

\boldparagraph{Binary Systems.}\label{binary systems} For a given $(q_0, p_0)$ the momentum 
$(\alpha_0, u_0) = J(q_0, p_0)$ can be calculated directly
and the isotropy group $G_{(q_0, p_0)}$ in the form appearing in the Table can be easily determined using \ref{Apendice A}. After this, by inspection of Table we can find the uniquely determined column having those data in the first and second row. This column contains basic information of the reduction process.

The problem of describing the motion of two point masses subjected to gravitational forces  has been solved by Newton’s Laws which can be written in Hamiltonian form where the Hamiltonian $H_N$ defined on $M$ is given by kinetic plus potential energy. However, Newton’s Laws give only an approximation to the motion determined by Einstein’s gravitational field in empty space which is more consistent with astronomical observation. In \cite{Da2012} and \cite{Schafer} (see Section 6, page 47), among several other references, a method based on Post-Newtonian 
Hamiltonians $H_N\equiv H_{0PN}, H_{1PN}, H_{2PN},
H_{3PN}, H_{4PN}, ...$,  is developed. Each Post-Newtonian Hamiltonian is defined on  $M$ and is invariant under the action of $SE(3)$, therefore Theorem \ref{Teorema principal} can be readily applied. To he best of our knowledge the methods of optimal reduction have not been systematically applied in the literature to this important physical example. 

\section{Appendix}

\subsection{Appendix A}\label{Apendice A}{\bf Isotropy subgroups of the action $T^*\phi$.}

The set $\bold{Is}$ of all isotropy subgroups of the action $T^*\phi$ can be classified in $13$ disjoint classes. In order to prove this, we will describe in detail all the possible cases of isotropy subgroups for the action $T^*\phi$ and we will define $13$ disjoint cases so each point of $M$ belongs to one and only one of them.  

\vspace{.2cm}
{\it $(a_1)$}: $m_{1}=(q^1,q^2,0,0)$,\ $q^1-q^2=0$; $SE(3)_{m_{1}}=G^1(q^1)=G^1(q^2)\simeq SO(3)$.

{\it $(a_2)$}: $m_{2}=(q^1,q^2,0,0)$,\ $q^1\parallel q^2$ , $q^1\neq q^2$; $SE(3)_{m_{2}}=G^2(q^1-q^2)\simeq SO(2)$.

{\it $(a_3)$}: $m_{3}=(q^1,q^2,0,0)$,\ $q^1\nparallel q^2$; $SE(3)_{m_{3}}=G^3(q^1,q^1-q^2)=G^3(q^2,q^1-q^2)\simeq SO(2)$.

{\it $(a_4)$}: $m_{4}=(q^1,q^2,0,p^2)$,\ $p^2\neq 0$,\ 
$q^1-q^2\nparallel p^2$; $SE(3)_{m_{4}}=\{(I,0)\}=G^4$.

{\it $(a_5)$}: $m_{5}=(q^1,q^2,0,p^2)$,\ $p^2\neq 0$,\ 
$q^1-q^2\parallel p^2$,\ $q^1$ and $q^2$ not parallel to $p^2$;
$SE(3)_{m_{5}}=G^3(q^1,p^2)=G^3(q^2,p^2)\simeq SO(2)$.

{\it $(a_6)$}: $m_{6}=(q^1,q^2,0,p^2)$,\ $p^2\neq 0$,
$q^1\parallel p^2$,\ $q^2\parallel p^2$; $ SE(3)_{m_{6}}=G^2(p^2)\simeq SO(2)$.

{\it $(a_7)$}: $m_{7}=(q^1,q^2,p^1,0)$,\ $p^1\neq 0$,\
$q^1-q^2\nparallel p^1$; $SE(3)_{m_{7}}=\{(I,0)\}=G^4$.

{\it $(a_8)$}: $m_{8}=(q^1,q^2,p^1,0)$,\ $p^1\neq 0$,\
$q^1-q^2\parallel p^1$ ,\ $q^1$ and $q^2$ not parallel to $p^1$;
$SE(3)_{m_{8}}=G^3(q^1,p^1)=G^3(q^2,p^1)\simeq SO(2)$.

{\it $(a_9)$}: $m_{9}=(q^1,q^2,p^1,0)$,\ $p^1\neq 0$,\ 
$q^1\parallel p^1$,\ $q^2\parallel p^1$; $SE(3)_{m_{9}}=G^2(p^1)\simeq SO(2)$.

{\it $(a_{10})$}: $m_{10}=(q^1,q^2,p^1,p^2)$,\ $p^1\parallel p^2$,\ $p^1\neq 0$, $p^2\neq 0$, $q^1-q^2\nparallel p^1$; 
$SE(3)_{m_{10}}=\{(I,0)\}=G^4$.

{\it $(a_{11})$}: $m_{11}=(q^1,q^2,p^1,p^2)$,\ $p^1\parallel  p^2$,\ $p^1\neq 0$, $p^2\neq 0$,\ $q^1-q^2\parallel p^1$,\ $q^1$ and $q^2$ not parallel to $p^1$; $SE(3)_{m_{11}}=G^3(q^1,p^1)=G^3(q^2,p^1)\simeq SO(2)$.

{\it $(a_{12})$}: $m_{12}=(q^1,q^2,p^1,p^2)$,\ $p^1\parallel  p^2$,\ $p^1\neq 0$,\ $p^2\neq 0$,\ $q^1$ and $q^2$
parallel to $p^1$; $ SE(3)_{m_{12}}=G^2(p^1)\simeq SO(2)$.

{\it $(a_{13})$}: $m_{13}=(q^1,q^2,p^1,p^2)$,\ $p^1\nparallel p^2$; $SE(3)_{m_{13}}=\{(I,0)\}=G^4$.

We shall omit the details of the proof.
\ \\

\subsection{Appendix B}\label{Apendice B}{\bf Proyective-cylindrical coordinates.}

The projectivization $\mathcal{P}(V)$ of a vector space $V$ is by definition the manifold of all (straight) lines of $V$ containing the origin. Let $x = x_1 {\bf e}_1 +  x_2 {\bf e}_2 +x_3 {\bf e}_ 3$, where  ${\bf e}_1, {\bf e}_2, {\bf e}_3$ is a positively oriented orthonormal basis in $\mathbb{R}^3$. For each $x$ such that $(x^1)^2 + (x^2)^2 >0$  we will denote $(z,s,x^3)$ the projective-cylindrical coordinates of $x$, where, by definition, $z\in \mathbb{R}$, $s\in [0, \pi)$, $x^1 = z\cos(s)$, $x^2 = z\sin(s)$. We will call $\ell_s \in \mathbb{R}P^1$ the line in the plane generated by $\{{\bf e}_1, {\bf e}_2\}$, forming an angle $s\in [0, \pi)$ with respect to the axis ${\bf e}_1$, thus $\mathbb{R}P^1$ denotes the projectivization of the plane perpendicular to $\bold{e}_3$. Therefore $\beta_s:=\cos(s){\bf e}_1+\sin(s) {\bf e}_2$ belongs to $\ell_s$. 

We will denote $\Pi_s$ the plane containing $\beta_s$ and ${\bf e}_3$. Then $\mathcal{P}(\Pi_s)$ is the 1-dimensional projective space of all lines of $\Pi_s$ containing the origin $\{0\}$. For instance $\ell_s$ and the line containing $\bold{e}_3$ belong to $P(\Pi_s)$. Elements of $\mathcal{P}(\Pi_s)$ will be usually denoted $\ell_{\nu}^{\Pi_s}$ where $\nu \in [0,\pi)$ is the angle between $\beta_s$ and $\ell_{\nu}^{\Pi_s}$,
and the orientation of $\Pi_s$ is given by the order $(\beta_s, \bold{e}_3)$. We will denote $\varphi^{\Pi_0}: \mathcal{P}(\Pi_0) \times\mathbb{R}^+ \times \mathbb{R}\times \mathbb{R} 
\longrightarrow \mathbb{R}P^1 \times\mathbb{R}^+ \times 
\mathbb{R}\times \mathbb{R}$ the map given by $\varphi^ {\Pi_0} (\ell^ {\Pi_0}_{\nu}, r, \mu, \lambda) = (\ell_{\nu}, r, \mu, \lambda)$. Let $u_0 = |u_0|{\bf e}_3 \neq 0$, then every element $x$ of $\Pi_s$ can be written uniquely as $x = z\beta_s + \mu u_0$.
For each $s\in [0, \pi)$ there is a linear map $f_s:\Pi_0 \rightarrow \Pi_s$ given by $f_s (z\beta_0 + \mu u_0) =z\beta_s + \mu u_0$. In the limit $s\rightarrow 
\pi$, $f_s$ is transformed by continuity into $f_{\pi}: \Pi_0 \rightarrow \Pi_0$ given by $f_{\pi} (z\beta_0 + \mu u_0) := -z\beta_0 + \mu u_0$, which reverses the orientation of $\Pi_0$. By an abuse of notation we will sometimes denote
$f_{\pi}(q^1, q^2, p^1, p^2) = \left(f_{\pi}(q^1), f_{\pi}(q^2), f_{\pi}(p^1), f_{\pi}(p^2)\right)$.

If $(q,p)=(q^1, q^2, p^1, p^2)$ the fact that  $q^1$, $q^2$, $p^1$, $p^2 \in S$, where $S$ is a certain subset, will be sometimes written, using an abuse of notation, $(q,p) \in S$. We will denote ${\bf p}: S^1 \rightarrow  \mathbb{R}P^1$ the diffeomorphism given by ${\bf p}(e^{i2s}) = \ell_s$. 
For $\tau \in \mathbb{R}$ let $n_{\tau} \in \mathbb{Z}$ be defined by the condition $n_{\tau}\pi \leq \tau < (n_{\tau}+ 1)\pi$ and define $[\tau]_{\pi} = \tau – n_{\tau}\pi$. 

The main motivation for introducing projective cylindrical coordinates in this paper is that, in \emph{Case 7}, the flow of an invariant Hamiltonian preserves the sheaf of all planes $\Pi_s$. For instance, the flow of the Hamiltonian 
$<(q^1-q^2),p^1\times p^2>$ gives a rotation of planes 
$\Pi_s$ with angular velocity $u_0$. In the other hand, the flow of Hamiltonians of the type $\displaystyle\frac{(p^1)^2}{2m^1}+\displaystyle\frac{(p^2)^2}{2m^2}+f(\vert q^1-q^2\vert)$ leave each $\Pi_s$ fixed.

\subsection{Appendix C}\label{Apendice C}{\bf An invariant atlas of $\mathcal{J}^{-1}({\rho^7_{(0,u_0)}})$ and the reduced atlas on $M_{\rho^7_{(0,u_0)}}$.}

We shall use the notation introduced in \emph{Case 7} of Theorem \ref{Teorema principal}. Let
\begin{equation*}\begin{array}{rcl}
S_{\mathfrak{p}}&:=&\{(q,p)\in\mathcal{J}^{-1}(\rho^7_{(0,u_0)}):r_p=0\},\\
S_{\mathfrak{q}}&:=&\{(q,p)\in\mathcal{J}^{-1}(\rho^7_{(0,u_0)}):r_{q^1q^2}=0\}.\\
\end{array}\end{equation*}

It can be deduced using formula (\ref{optimo caso 7}) that $S_{\mathfrak{p}}$ is a submanifold of $\mathcal{J}^{-1}(\rho^7_{(0,u_0)})$ defined regularly by the equation 
$r_p = 0$ and, besides, that for each  $s\in[0,\pi)$, $S_{\mathfrak{p}}\cap \Pi_s$ has two connected components defined by the conditions $r_{q^1q^2}>0$ and $r_{q^1q^2}<0$. Using techniques similar to the ones used in \emph{Case 7} to prove that $J^{-1} (0, u_0)\cap M_{G^4}$ is arc connected, one
can prove that each point of $S_{\mathfrak{p}}$ can be joined by a continuous curve contained in $S_{\mathfrak{p}}$ to one of the connected components of $S_{\mathfrak{p}} \cap \Pi_0$. On the other hand, as in \emph{Case 7}, one can choose a continuous curve $(q(s), p(s)) = P^7_{(0,u_0)}(\ell_s, e^{i\tau}, r, \mu_{q^1}, \mu_{q^2},\lambda)$ in such a way that $r_{p(0)} = 0$ and $r_{q^1(0)q^2(0)} >0$ which implies that this curve is contained in $S_{\mathfrak{p}}$ and, besides, $r_{f_{\pi}(q^1(0))f_{\pi}(q^2(0))} = -r_{q^1(0)q^2(0)} <0$, so the points $(q(0), p(0))$ and $(f_{\pi}(q(0)), f_{\pi}(p(0)))$ belong to different connected components of of $S_{\mathfrak{p}} \cap \Pi_0$. We have proven that   $S_{\mathfrak{p}}$ is an arc-connected submanifold of $\mathcal{J}^{-1}(\rho^7_{(0,u_0)})$.
 
Similarly, $S_{\mathfrak{q}}$ is a submanifold of 
$J^{-1} (0, u_0)\cap M_{G^4}$ defined regularly by the equation 
$r_{q^1 q^2} = 0$ and, besides, for each  $s\in[0,\pi)$, $S_{\mathfrak{q}}\cap\Pi_s$ has two connected components defined by the conditions $r_p>0$ and $r_p<0$ but $S_{\mathfrak{q}}$ is a connected submanifold of $\mathcal{J}^{-1}(\rho^7_{(0,u_0)})$. We observe that $S_{\mathfrak{p}}\cap S_{\mathfrak{q}}=\emptyset$.
\ \\

We are going to define an atlas of the manifold $\mathcal{J}^{-1}(\rho^7_{(0,u_0)})$ by two charts, namely
\begin{equation*}\begin{array}{rcl}
W_{\mathfrak{p}}&:=&\mathcal{J}^{-1}(\rho^7_{(0,u_0)})-S_{\mathfrak{p}},\\
W_{\mathfrak{q}}&:=&\mathcal{J}^{-1}(\rho^7_{(0,u_0)})-S_{\mathfrak{q}}.\\
\end{array}\end{equation*}

Using an argument similar to the one that we used with $S_{\mathfrak{p}}$ we can show that for each $s\in[0,\pi)$, $W_{\mathfrak{p}}\cap\Pi_s$ has two connected components defined by $r_p >0$ and $r_p <0$ but $W_{\mathfrak{p}}$ is connected. Similarly, for each $s\in[0,\pi)$, $W_{\mathfrak{q}}\cap\Pi_s$ has two connected components defined by $r_{q^1 q^2} >0$ and 
$r_{q^1 q^2} < 0$ but $W_{\mathfrak{q}}$ is connected.

$W_{\mathfrak{p}}$ and $W_{\mathfrak{q}}$ are the image under $P^7_{(0,u_0)}$ of $\mathbb{R}P^1\times (S^1 - \{e^{i0}, e^{i\pi}\})\times \mathbb{R}^+ \times \mathbb{R}\times \mathbb{R}\times \mathbb{R}$ and $\mathbb{R}P^1\times (S^1 - \{e^{i\frac{\pi}{2}}, e^{i\frac{3\pi}{2}}\})\times \mathbb{R}^+ \times \mathbb{R}\times \mathbb{R}\times \mathbb{R}$, respectively.

We observe that each one of these charts is invariant under the action of $G_{\rho^7_{(0,u_0)}}$, so they become trivial principal bundles whose projections are the restriction of
$\pi_{\rho^7_{(0,u_0)}}$ to $W_{\mathfrak{p}}$ and $W_{\mathfrak{q}}$. The restriction of the diffeomorphism
$({\bf p}^7_{(0,u_0)})^{-1}$ to the orbit spaces
$\pi_{\rho^7_{(0,u_0)}} (W_{\mathfrak{p}})$  and 
$\pi_{\rho^7_{(0,u_0)}} (W_{\mathfrak{q}})$ are diffeomorphisms onto $\mathcal{P}(\Pi_0) - \{\ell^{\Pi_0}_0\}$ and
$\mathcal{P}(\Pi_0) - \{\ell^{\Pi_0}_{\frac{\pi}{2}}\}$ respectively. It follows that the restriction of the trivialization $\bold{P}^7_{(0,u_0)}$ of \emph{Case 7} to $G_{\rho^7_{(0,u_0)}} \times 
\pi_{\rho^7_{(0,u_0)}}(W_{\mathfrak{p}})$ and to $G_{\rho^7_{(0,u_0)}} \times\pi_{\rho^7_{(0,u_0)}}(W_{\mathfrak{q}})$ gives a trivialization of $W_{\mathfrak{p}}$ and $W_{\mathfrak{q}}$, respectively. Therefore we have the principal bundle isomorphisms $\bold{P}^7_{(0,u_0)}\circ [Id_{G_{\rho^7_{(0, u_0)}}}\times (\bold{p}^7_{(0,u_0)}\vert_{\mathcal{P}(\Pi_0)-\{\ell_0^{\Pi_0}\}})]:G_{\rho^7_{(0, u_0)}}\times (\mathcal{P}(\Pi_0)-\{\ell_0^{\Pi_0}\})\longrightarrow W_{\mathfrak{p}}$ and $\bold{P}^7_{(0,u_0)}\circ [Id_{G_{\rho^7_{(0, u_0)}}}\times (\bold{p}^7_{(0,u_0)}\vert_{\mathcal{P}(\Pi_0)-\{\ell_{\frac{\pi}{2}}^{\Pi_0}\}})]:G_{\rho^7_{(0, u_0)}}\times (\mathcal{P}(\Pi_0)-\{\ell_{\frac{\pi}{2}}^{\Pi_0}\})\longrightarrow W_{\mathfrak{q}}$  
\ \\


\begin{thebibliography}{99}                                                                                               %

\bibitem[AM78]{AM}
Ralph Abraham and Jerrold ~E. Marsden, \emph{Foundations of mechanics},
  Benjamin/Cummings Publishing Co. Inc. Advanced Book Program, Reading, Mass.,
  1978, Second edition, revised and enlarged, With the assistance of Tudor
  Ratiu and Richard Cushman.

\bibitem[Arn66]{Arnold2}
Vladimir I.~Arnold, \emph{Sur la g\'eom\'etrie diff\'erentielle des groupes de {L}ie de
  dimension infinie et ses applications \`a l'hydrodynamique des fluides
  parfaits}, Ann. Inst. Fourier (Grenoble) \textbf{16} (1966), no.~fasc. 1,
  319--361.
  
\bibitem[CHMR98]{LREPE}
Hern{\'a}n Cendra, Darryl~D. Holm, Jerrold~E. Marsden, and Tudor~S. Ratiu,
  \emph{Lagrangian reduction, the {E}uler-{P}oincar\'e equations, and
  semidirect products}, Geometry of differential equations, Amer. Math. Soc.
  Transl. Ser. 2, vol. 186, Amer. Math. Soc., Providence, RI, 1998, pp.~1--25.

\bibitem[CMR01]{LRBS}
Hern{\'a}n Cendra, Jerrold~E. Marsden, and Tudor~S. Ratiu, \emph{Lagrangian
  reduction by stages}, Mem. Amer. Math. Soc. \textbf{152} (2001), no.~722,
  x+108.


\bibitem[Da2012]{Da2012}
Thibault Damour, \emph{The General Relativistic Two Body Problem and the Effective One Body Formalism}, Proceedings of the Conference "Relativity and Gravitation - 100 years after Einstein in Prague", Prague, 25-29 June 2012. arXiv:1212.3169

\bibitem[MdL]{MDL89}
Manuel de León, Paulo R. Rodriguez, \emph{Methods of Differential Geometry in Analytical Mechanics}, North Holland, 1989, Volume 158 1st Edition.

\bibitem[Gol51]{Goldstein}
Herbert Goldstein, \emph{Classical {M}echanics}, Addison-Wesley Press, Inc.,
  Cambridge, Mass., 1951.

\bibitem[GS77]{guillsternbrg}
Victor Guillemin and Shlomo Sternberg, \emph{Geometric asymptotics}, American
  Mathematical Society, Providence, R.I., 1977, Mathematical Surveys, No. 14.
  
\bibitem[GS80]{guillsternbrg3}
Victor Guillemin and Shlomo Sternberg, \emph{The moment map and collective motion}, Ann. Physics \textbf{127}
  (1980), no.~1, 220--253.

\bibitem[GS90]{guillsternbrg2}
Victor Guillemin and Shlomo Sternberg, \emph{Symplectic techniques in physics}, second ed., Cambridge
  University Press, Cambridge, 1990.

\bibitem[GS05]{guillsternbrg4}
Victor Guillemin and Shlomo Sternberg, \emph{The moment map revisited}, J. Differential Geom. \textbf{69}
  (2005), no.~1, 137--162.

\bibitem[Ham34]{Hamilton}
William.~R. Hamilton, \emph{On a general method in dynamics}, Phil. Trans. Roy.
  Soc. Lon. (1834), 247--308.

\bibitem[Jac66]{Jacobi}
Carl ~G. ~J. Jacobi, \emph{Vorlesungen \"{u}uber dynamik}, Based on lectures given in
  1842-3. Verlag G. Reimer. Reprinted by Chelsea, 1969 (1866).

\bibitem[JS98]{Saletan}
Jorge~V. Jos{\'e} and Eugene~J. Saletan, \emph{Classical dynamics}, Cambridge
  University Press, Cambridge, 1998, A contemporary approach.


\bibitem[Lag88]{Lagrange}
Joseph ~L. Lagrange, \emph{Mécanique analytique}, Chez la Veuve Desaint, Paris.
  (1788).

\bibitem[MMO07]{HRBS}
Jerrold~E. Marsden, Gerard Misiolek, Juan Pablo Ortega, Matthew Perlmutter and
  Tudor~S. Ratiu, \emph{Hamiltonian reduction by stages}, Lecture Notes in
  Mathematics, vol. 1913, Springer, Berlin, 2007.
  
\bibitem[MR86]{MR-reduction}
Jerrold~E. Marsden and Tudor~S. Ratiu, \emph{Reduction of {P}oisson manifolds},
  Lett. Math. Phys. \textbf{11} (1986), no.~2, 161--169.

\bibitem[MR94]{MR}
Jerrold~E. Marsden and Tudor~S. Ratiu, \emph{Introduction to mechanics and
  symmetry}, Texts in Applied Mathematics, vol.~17, Springer-Verlag, New York,
  1994, A basic exposition of classical mechanical systems.


\bibitem[MW01]{MW-coments}
Jerrold~E. Marsden and Alan Weinstein, \emph{Comments on the history, theory,
  and applications of symplectic reduction}, Quantization of singular
  symplectic quotients, Progr. Math., vol. 198, Birkh\"auser, Basel, 2001,
  pp.~1--19.

\bibitem[OR02]{O-R}
Juan Pablo Ortega and Tudor~S. Ratiu, \emph{The optimal momentum map},
  Geometry, mechanics, and dynamics, Springer, New York, 2002, pp.~329--362.

\bibitem[OR04]{O-R_libro}
Juan Pablo Ortega and Tudor~S. Ratiu, \emph{Momentum maps and hamiltonian reduction}, Progress in
  Mathematics, vol. 222, Birkh\"auser Boston Inc., Boston, MA, 2004.

\bibitem[Poi92]{Poincare1}
Henri ~Poincaré, \emph{Les formes d'équilibre d'une masse fluide en rotation},
  Revue Générale des Sciences \textbf{3} (1892), 809--815.

\bibitem[Poi01]{Poincare2}
Henri ~Poincaré, \emph{Sur une forme nouvelle des équations de la mécanique}, C.R.
  Acad.Sci. \textbf{132} (1901), 369--371.

\bibitem[S18]{Schafer}
G. Schäfer, and P. Jaranowski, \emph{Hamiltonian formulation of general relativity and Post-Newtonian dynamics of compact binaries}, Living Rev Relativ 21, 7 (2018). https://doi.org/10.1007/s41114-018-0016-5.

\bibitem[Sma70a]{Smale1}
Stephen Smale, \emph{Topology and mechanics. {I}}, Invent. Math. \textbf{10} (1970),
  305--331.

\bibitem[Sma70b]{Smale2}
Stephen Smale, \emph{Topology and mechanics. {II}. the planar {$n$}-body problem},
  Invent. Math. \textbf{11} (1970), 45--64.



\end{thebibliography}
\end{document}